\documentclass[a4paper,11pt]{article}
\pdfoutput=1 

\usepackage{jcappub} 
\usepackage[T1]{fontenc} 

\usepackage{natbib}
\usepackage{cancel}
\usepackage{enumitem}    
\usepackage{amsthm}         
\usepackage{amsmath} 	   
\usepackage{amssymb}       
\usepackage{amsfonts}
\usepackage{graphicx} 
\usepackage{dblfloatfix}	    
\usepackage{verbatim}
\usepackage{epstopdf}
\usepackage{tensor}
\usepackage{epsfig,multicol,bbm}
\usepackage{color}
\usepackage{colortbl}
\usepackage{float}
\usepackage{multirow}
\usepackage{array}

\usepackage{subfig}

\allowdisplaybreaks

\title{Ultralocality and Slow Contraction}
\author[a,b,1]{Anna Ijjas,}
\note{Corresponding Author}
\author[c]{Andrew P. Sullivan,}
\author[c]{Frans Pretorius,}
\author[c]{Paul J. Steinhardt}
\author[d]{and William G. Cook}

\affiliation[a]{Max Planck Institute for Gravitational Physics (Albert Einstein Institute), Hannover, 30167, Germany}
\affiliation[b]{Institute for Gravitational Physics, Leibniz University, Hannover, 30167, Germany}
\affiliation[c]{Department of Physics, Princeton University, Princeton, NJ, 08544, USA}
\affiliation[d]{Theoretisch-Physikalisches Institut, Friedrich-Schiller-Universit\"at, Jena, 07743, Germany}

\emailAdd{anna.ijjas@aei.mpg.de}

\abstract{
We study the detailed process by which slow contraction smooths and flattens the universe using an improved numerical relativity code that accepts initial conditions with non-perturbative 
deviations from homogeneity and isotropy along two independent spatial directions.   
Contrary to common descriptions of the early universe,
we find that the geometry first rapidly converges to an inhomogeneous, spatially-curved and anisotropic {\it ultralocal} state in which all spatial 
gradient contributions to the equations of motion decrease as an exponential in time to negligible values. This is followed by a second stage in which the geometry converges to a homogeneous, spatially flat and isotropic spacetime. In particular,  the decay appears to follow the same history whether the entire spacetime or only parts of it are smoothed by the  end of slow contraction.}


\begin{document}
\maketitle 
\raggedbottom

\section{Introduction}
\label{sec_intro}

Explaining the observed uniformity of the universe on large scales is one of the longest-standing challenges in physical cosmology.  Although observations show that the evolution of the universe since the onset of radiation domination is well-described by the laws of general relativity, this is only possible for a particular set of initial conditions. For almost all other initial conditions, the universe would have evolved towards an inhomogeneous, anisotropic and spatially curved geometry. In this sense, the large-scale properties of our universe appear to be special.

In general, initial conditions are independent of the dynamical evolution equations. It is therefore remarkable that the  cosmic initial conditions problem and the possible solutions to it can be directly related to a basic feature of the Einstein field equations:  namely, the characteristic physical length scale is in general different from and evolves differently with time than the characteristic length scale of interactions \cite{Ijjas:2018qbo}.
In particular, in a Friedmann-Robertson-Walker (FRW) space-time (like our large-scale universe), which is given by the line element,
\begin{equation}
\label{FRW-line-el}
{\rm d}s^2 = - {\rm d}\tau^2 + a^2(\tau)\delta^{ij}{\rm d}x_i{\rm d}x_j\,,
\end{equation}
physical distances evolve as the scale factor $a(\tau)$. The characteristic length scale of interactions, on the other hand, is given by the Hubble radius $|H^{-1}|$, where $H\equiv d\ln a/d\tau$. 
Today the volume that encompasses the regions of spacetime which have been in causal contact has a radius 
roughly equal to $1/H_0$, where $H_0$ is the current Hubble parameter.
(Throughout, quantities are given in reduced Planck units and the scale factor is normalized such that $a(\tau_{\rm i})=1$ at some initial time $\tau_{\rm i}$.)

From the Einstein equations for an FRW space-time, one obtains the relation 
\begin{equation}
\label{Friedmann-const-simple}
|H^{-1}| \propto a^{\varepsilon}\,,
\end{equation}
from which it is immediately obvious that the Hubble radius evolves at a different rate than the scale factor and the relative growth rate is determined by the equation of state 
\begin{equation}
\label{e.o.s.}
\varepsilon\equiv \frac32 \left(1+ \frac{p}{\varrho}\right),
\end{equation}
where $p$ is the pressure and $\varrho$ is the energy density of the dominant stress-energy component. For example, in a radiation ($\varepsilon = 2$) or matter ($\varepsilon = 3/2$) dominated universe, the Hubble radius grows faster than the scale factor. As a result, the volume comprising the observable universe today extrapolated back to the onset of radiation domination contained  approximately $e^{180} \sim 10^{80}$ causally independent Hubble-sized patches at the onset of radiation domination.
Consequently, the uniformity observed today would require some mechanism to smooth and synchronize those  $10^{80}$  patches by the onset of radiation domination. 

Classical smoothing mechanisms, inflation \cite{Guth:1980zm,Albrecht:1982wi,Linde:1981mu} and slow contraction \cite{Khoury:2001bz}, rely on a simple yet elegant idea to achieve this smoothing and synchronization: 
by causing a single initial homogeneous and isotropic Hubble volume $|H_{\rm beg}^{-3}|$ to 
evolve to encompass exponentially many (at least $10^{80}$) Hubble volumes $|H_{\rm end}^{-3}|$ by the time the smoothing phase ends and the radiation dominated phase begins. After the 60 $e$-foldings of subsequent radiation and matter dominated decelerated expansion, it is a subvolume of the initial Hubble-sized patch that makes up the observable universe.  For example, slow contraction ($\varepsilon>3$) is a classical smoother because the scale factor shrinks much more slowly  than the Hubble radius as given in Eq.~\eqref{Friedmann-const-simple}.  For typical values of the equation of state ($\varepsilon \sim 50$), the initial Hubble radius ($|H_{\rm beg}|^{-1}$)  shrinks by a factor of $2^{50}$ while the scale factor (and, hence, the radius of the initial Hubble volume) decreases by only a factor of 2 \cite{Ijjas:2019pyf}.

Note, that, contrary to (Newtonian) intuition, in a universe where gravity follows the laws of General Relativity, both expansion and contraction can smooth or unsmooth the cosmological background. For example, decelerated expansion ($\varepsilon>1$) and fast contraction ($\varepsilon<3$) both amplify small deviations from homogeneity and isotropy. Accelerated expansion ($\varepsilon<1$) and slow contraction ($\varepsilon>3$), on the other hand, 
both suppress small deviations from homogeneity and isotropy.

The virtue of classical smoothing mechanisms is that they show how the observed features of the large-scale universe might be traced back to the features of a single (rather than $10^{26}$) initial Hubble-sized patch. Yet classical smoothing is {\it not} sufficient to solve the cosmic initial conditions problem for two reasons: 
First, classical smoothing does not guarantee stability to quantum fluctuations around the classical background.  For example, despite  being a classical smoother, inflation famously suffers from a quantum runaway problem, leading to eternal inflation and the multiverse \cite{Steinhardt:1982kg,Vilenkin:1983xq,Guth:2007ng}.  
Second, classical smoothing relies on assuming that the initial Hubble volume ($|H_{\rm beg}|^{-3}$)  is {\it already} homogeneous  {\it before} the smoothing phase begins, which is clearly a very special initial condition as difficult to account for as  the problem one is trying to solve in the first place.
Either being unstable to quantum fluctuations or not being robust enough to 
smooth the universe for initial conditions that lie outside the perturbative regime of FRW space-times presents a roadblock to solving the cosmic initial conditions problem.

Currently, slow contraction is the only known classical smoothing mechanism that is both   {\it quantum stable}  and {\it robust} \cite{Cook:2020oaj,Ijjas:2020dws}. Notably, in Ref.~\cite{Ijjas:2020dws}, we observed signs that smoothing and flattening during slow contraction are achieved in a particular way: namely, independent of the initial data, the evolution first becomes ultralocal, {\it i.e.}, spatial gradients quickly become unimportant, even before homogeneity, flatness and isotropy are achieved. This is contrary to common descriptions of the early universe where it is assumed that smoothing  proceeds by first converging to a homogeneous spacetime that is in general spatially-curved curved and anisotropic.

Ultralocal behavior during contraction has been considered for several decades in different contexts without reference to smoothing and flattening. Originally, it was conjectured in Ref.~\cite{Belinsky:1970ew} that, in contracting vacuum space-times, spatial gradients, measured relative to parallel transported coordinates, are `velocity dominated,' {\it i.e.} spatial gradients in the equations of motion become small compared to the time derivatives. Several numerical analyses studying how relativistic space-times approach a putative singularity provide evidence for the conjecture in some special settings, assuming certain symmetry conditions or a particular matter source (vacuum, stiff fluid, or a free scalar) \cite{Berger:1998vxa,Lim:2009dg,Garfinkle:2020lhb}.  Mathematically, the conjecture is not (yet) proven in the global setting. In the special cases where a rigorous proof could already be obtained, ultralocality is understood as following from the stability  of Kasner spacetimes \cite{Andersson:2000cv,Damour:2002tc,Rodnianski:2018hin}.

In this paper, we go beyond Ref.~\cite{Ijjas:2020dws} 
using an improved numerical relativity code that accepts initial conditions with non-perturbative deviations from homogeneity and isotropy along two independent spatial directions to do a more extensive, detailed
study.   
We find that in relativistic spacetimes where matter is sourced by a minimally coupled, ordinary scalar field with negative potential energy density, the smoothing during slow contraction  occurs {\it in general} through first converging to an ultralocal state. To demonstrate this effect, we numerically solve the Einstein-scalar field equations in generic non-perturbative and non-symmetric settings, confirming and generalizing the results obtained in Ref.~\cite{Ijjas:2020dws}. As one of the highlights, we show that spatial points in regions that eventually become smooth and flat explore the same dynamical history  whether the entire spacetime or only parts of it end up smooth and flat by the end of the smoothing phase. In addition, we demonstrate that ultralocality is always achieved in a particular way where the gradient fall-off follows an exponential behavior in time. 

\section{Numerical evolution scheme}
\label{sec_scheme}

To carry out our non-perturbative, numerical calculations, we shall employ the orthonormal tetrad formulation of the Einstein-scalar field equations. A comprehensive introduction to the formulation including the derivation of the partial differential equation system is given in Ref.~\cite{Ijjas:2020dws}. Here, we will not repeat the same details but provide a complementary yet self-contained overview which underlies the results in this paper.

\subsection{Geometric variables}
\label{subsec:geomvar}

Tetrad formulations have in common that they {\it locally} represent each space-time point by a family of unit basis four-vectors $\{e_0,e_1, e_2, e_3 \}$ rather than (scalar) coordinates.  Here, the timelike {\it vierbein} $e_0$ defines the future directed timelike congruence, to which it is tangent. The spacelike unit four-vectors $\{e_1, e_2, e_3 \}$ span a spatial triad, each lying in a rest three-space of $e_0$. 
The basis four-vectors define a local Lorentz {\it frame} with the spacetime metric being given by the inner product "$\cdot$" of the {\it vierbein}. 
For an {\it orthonormal} tetrad, 
\begin{equation}
\label{tetrad-metric}
g_{\alpha\beta} \equiv e_{\alpha} \cdot e_{\beta} = \eta_{\alpha\beta}\,,
\end{equation}
where $\eta_{\alpha\beta} = {\rm diag}(-1, 1,1,1)$ is the Minkowski metric. Tetrad frame indices are raised and lowered with $\eta_{\alpha\beta}$.

The forty geometric {\it variables} of the formulation are the sixteen tetrad vector components $\{e_{\alpha}{}^{\mu}\}$ and the twenty-four Ricci rotation coefficients,
\begin{equation}
\gamma_{\alpha\beta\lambda} \equiv e_{\alpha} \cdot \nabla_{\lambda} \,e_{\beta}\,, 
\end{equation}
where $\nabla_{\lambda} \equiv e_{\lambda}{}^{\mu}\nabla_{\mu}$ is the spacetime covariant derivative projected onto the {\it vierbein} $e_{\lambda}$.
Throughout, spacetime indices $(0-3)$ are Greek and spatial indices are Latin $(1-3)$. The beginning of the alphabet ($\alpha, \beta, \gamma$ or $a, b, c$) denotes tetrad indices and the middle of the alphabet ($\mu, \nu, \rho$ or $i, j, k$) denotes coordinate indices.

The Ricci rotation coefficients define the deformation of the tetrad frame $\{e_0,e_1, e_2, e_3 \}$ when moving from point to point.
The role of these geometric quantities becomes apparent by performing a (1+3) split relative to the timelike congruence tangent to $e_0$: since $\gamma_{\alpha\beta\lambda}$ is antisymmetric in its first two indices,  the Ricci rotation coefficients that have at least one timelike index can be described through fifteen three-dimensional quantities 
\begin{eqnarray}
 \gamma_{a00} &=& - \gamma_{0a0} \equiv b_a,\\
 \gamma_{ab0} &=& - \gamma_{ba0} \equiv \epsilon_{abc}\Omega^c ,\\
 \gamma_{0ab} &=& - \gamma_{a0b} \equiv -K_{ba} ,
\,,
\end{eqnarray}
where $\epsilon_{abc}$ is the Levita symbol; and the  rotation coefficients with purely spatial indices $\gamma_{abc}$ are described by the nine three-tensor components
\begin{equation}
N_{ab}\equiv {\textstyle \frac12} \epsilon_b{}^{cd}  \gamma_{cda}\,.
\end{equation}
The three-vectors $b_a$ and $\Omega_a$ are frame {\it gauge} quantities, defining the local proper acceleration and the local angular velocity of the spacelike triad relative to Fermi propagated axes, respectively. 
The eighteen dynamical variables are comprised by the components of the shear (or rate-of-strain) tensor $K_{ab}$ and the components of  the induced curvature tensor $N_{ab}$ associated with the spatial three-congruence.

Here, we require that 
\begin{itemize}
\item[-] the spatial triad is `Fermi propagated' ($\Omega_a \equiv 0$), meaning that it is a local, inertially non-rotating frame; and
\item[-] the timelike congruence is hypersurface orthogonal ($K_{ab}\equiv K_{ba}$), meaning that the timelike {\it vierbein} is the future directed unit normal to the spacelike hyersurfaces of constant time $\{\Sigma_t\}$ with the spatial tetrads being tangent to $\{\Sigma_t\}$.
\end{itemize}
Note that, with this frame gauge choice, 
the acceleration of the congruence $b_a$ is given through
\begin{equation}
b_a \times e_0(x^0) = - e_a\, e_0(x^0)\,,
\end{equation}
where $x^0$ is the time coordinate of $\Sigma_t$,
and the dynamical variables obtain definite physical meaning: the three-tensor $K_{ab}$ describes the extrinsic curvature of the constant time hypersurfaces $\{\Sigma_t\}$ while the components of the three-tensor $N_{ab}$ are the spatial (or intrinsic) curvature variables. Furthermore, all three-tensor components $\{K_{ab}, N_{ab}\}$ act as scalars on $\{\Sigma_t\}$. (Throughout, $\times$ denotes scalar-scalar multiplication.)

Of course, the geometric variables must be supplemented by the dynamical variables describing the matter that we will specify next.

\subsection{Matter source}

As summarized in the Introduction, classical smoothing through slow contraction is based on the idea that a stress-energy source which behaves as a perfect fluid with super-stiff equation of state ($\varepsilon > 3$) breaks up the initial Hubble volume into $\sim10^{80}$ self-similar homogeneous, isotropic and flat Hubble patches at the end of the contracting phase. 

A standardly used  microphysical model that can generate a period of slow contraction has  a stress-energy consisting of
an ordinary scalar field $\phi$ minimally coupled to Einstein gravity with canonical kinetic energy and a negative potential $V(\phi)$.
Indeed, on a smooth and flat FRW background, a scalar field $\phi$ behaves like a perfect fluid with energy density and pressure defined as
\begin{eqnarray}
\label{rho-phi}
\varrho &\equiv& {\textstyle \frac12} \dot{\phi}^2 + V(\phi),\\
\label{p-phi}
p &\equiv& {\textstyle \frac12} \dot{\phi}^2 - V(\phi)\,,
\end{eqnarray}
where dot denotes differentiation with respect to the physical FRW time coordinate $\tau$,
such that the scalar field equation of state is given by
\begin{equation}
\varepsilon \equiv \frac32\left(1+ \frac{p}{\rho}\right) = 3\times \frac{{\textstyle \frac12} \dot{\phi}^2}{{\textstyle \frac12} \dot{\phi}^2+ V(\phi)}\,,
\end{equation}
and the Einstein-scalar field equations reduce to the Friedmann equations,
\begin{eqnarray}
\label{FRW-const}
3H^2 &=& \varrho = {\textstyle \frac12} \dot{\phi}^2 + V(\phi),\\
\label{FRW-eq}
- 2\dot{H} &=& \varrho + p =  \dot{\phi}^2.
\end{eqnarray}
In particular, for a negative exponential potential 
\begin{equation}
\label{V-def}
V(\phi) = V_0 \exp(-\phi/M)\,,
\end{equation}
where $M$ is the characteristic mass scale associated with the scalar field and $V_0<0$,
the Friedmann equations admit the scaling attractor solution 
\begin{equation}
\label{FRW-scaling-sol}
a(\tau) = (-\tau)^{1/\varepsilon},\quad \phi(\tau) =  \sqrt{\frac{2}{\varepsilon}} \times \ln \left(- \sqrt{V_0\,\frac{\varepsilon^2}{3-\varepsilon}}\times \tau \right),\quad \varepsilon = {\textstyle \frac12} M^{-2}.
\end{equation}
Note that we have chosen coordinates such that the physical time variable $\tau<0$ is running from large negative to small negative values during the slow contraction phase. For typical values of $M$, say $M \sim 0.1$ (in reduced Planck units) or  $\varepsilon \sim 50$,  the scale factor $a$ (and all physical distances) shrinks by only a factor of two or three during the entire slow contraction phase while the Hubble radius $|H^{-1}|$ decreases by a factor of $2^{50}$. 
 
 When testing for robustness to initial conditions, the question  is whether the non-linear Einstein-scalar system of coupled partial differential equations (PDEs),
 \begin{eqnarray}
 \label{Einstein-eq}
R_{\alpha\beta} &=& T_{\alpha\beta}  - {\textstyle \frac12}\eta_{\alpha\beta}T_{\lambda}{}^{\lambda} \,,\\
\label{KleinGordon}
\Box\phi &=& V_{,\phi}\,,
\end{eqnarray}
 where $R_{\alpha\beta}$ is the Ricci tensor and the stress-energy is given by
 \begin{equation}
 \label{stress-energy}
T_{\alpha\beta} \equiv \nabla_\alpha\phi \nabla_\beta\phi 
- \left( {\textstyle \frac12}  \nabla_{\lambda} \phi \nabla^{\lambda} \phi 
+ V(\phi) \right) \eta_{\alpha\beta}
  \,,
\end{equation}
 {\it generically} evolves towards the simple, homogeneous  Friedmann system~(\ref{FRW-const}-\ref{FRW-eq}) of ordinary differential equations(ODEs), especially in situations where the initial data lies far outside the perturbative regime of FRW spacetimes. 
  
 With the frame gauge choice as described in Sec.~\ref{subsec:geomvar}, the macroscopic matter variables take the following form:
 \begin{eqnarray}
 \varrho &\equiv& e_0{}^\alpha e_0{}^\beta T_{\alpha\beta} = {\textstyle \frac12} D_0 \phi D_0\phi  + {\textstyle \frac12} D_a \phi D^a\phi + V(\phi) 
 ,\\
j_a &\equiv&e_0{}^\alpha e_a{}^\beta T_{\alpha\beta}  =
- D_0\phi D_a\phi
,\\
 s_{ab} &\equiv& e_a{}^\alpha e_b{}^\beta T_{\alpha\beta}  = D_a \phi D_b\phi  + \left({\textstyle \frac12} D_0 \phi D_0\phi  - D_c \phi D^c\phi - V(\phi) \right)\delta_{ab}
 ,\\
p &\equiv& {\textstyle \frac13} s_a{}^a = {\textstyle \frac12} D_0 \phi D_0\phi  - {\textstyle \frac16} D_a \phi D^a\phi - V(\phi) 
 \,,
\end{eqnarray}
 where $\varrho$ is the energy density, $j_a$ the three-momentum flux, $s_{ab}$ the spatial stress tensor, and $p$ the pressure; $D_0$ denotes the Lie derivative along $e_0$ and $D_a$ is the directional derivative along $e_a$.
Note that, when gradients are non-negligible, $j_a, s_{ab}\neq 0$, a hypersurface-orthogonal tetrad frame gauge is  {\it not} the same as co-moving frame of the scalar matter field.

\subsection{Evolution and constraint equations in orthonormal tetrad form}

Numerical relativity simulations evolve variables specified on an initial spacelike hypersurface $\Sigma_{t_0}$  which are subject to a system of partial differential equations (PDEs). 
Accordingly, for the non-perturbative, numerical solution of the Einstein-scalar field equations~(\ref{Einstein-eq}-\ref{stress-energy}), we must represent the tetrad variables $\{ \gamma_{abc}, e_\alpha \}$ in terms of scalar functions that depend on coordinates; and we must also represent the directional derivatives along tetrad vectors $D_{\alpha}$ in terms of partial derivatives acting upon scalars which are functions of the coordinates. Finally, the coordinate gauge must be fixed such that, for appropriately defined initial data and boundary conditions, the resulting PDE system is {\it well-posed}, yielding a unique solution that continuously depends on the initial data.

\subsubsection{Coordinate representation of tetrad variables}
Having fixed the tetrad frame gauge to be Fermi propagated and hypersurface-orthogonal, the coordinate representation of the tetrad variables becomes particularly straightforward:
 the Ricci rotation coefficients which are true dynamical variables, namely the six components of the extrinsic curvature tensor $K_{ab}$ and the nine components of the intrinsic curvature tensor $N_{ab}$, act as scalar functions of coordinates on spatial hypersurfaces of constant time $\Sigma_t$. Hence, it remains to write the tetrad vector components $\{e_\alpha{}^\mu\}$ as coordinate functions.

First, we introduce the matrix $\{\lambda_{\alpha}{}^{\mu}\}$ that defines the transformation between tetrad and coordinate vectors,
 \begin{equation}
\label{trafo-matrix}
e_\alpha \equiv \lambda_{\alpha}{}^{\mu} e_\mu\,.
\end{equation}
 With $e_0$ being the future-directed timelike normal to the spacelike hypersurface of constant time $\Sigma_t$   and $e_a$ being tangent to $\Sigma_t$, the matrix elements $\{\lambda_{\alpha}{}^{\mu}\}$ can easily be identified with quantities of the 3+1 (coordinate-based) Arnowitt-Deser-Misner (ADM) formalism \cite{Arnowitt:1959ah}: 
\begin{equation}
\lambda_0{}^0 =  \frac{1}{N},\quad
\lambda_0{}^i = - \frac{N^i}{N}, \quad
\lambda_a{}^0 = 0,\quad
\lambda_a{}^i = E_a{}^i\,,
\end{equation}
where $N$ is the ADM lapse function  and $N^i$ the ADM shift vector, and
the coordinate metric,
\begin{equation}
g^{\mu\nu} = \eta^{\alpha\beta}\lambda_{\alpha}{}^{\mu}\lambda_{\beta}{}^{\nu}\,,
\end{equation}
 is given by
 \begin{equation}
g^{00}= - \frac{1}{N^2},\quad g^{0i}= -\frac{N^i}{N^2},\quad g^{ij}= E_a{}^i E_a{}^j\,.
\end{equation}
In particular, orthogonal hypersurface-slicing implies that the tetrad and coordinate lapse function and shift vector coincide. This is because, in this special frame gauge, the tetrad congruence simultaneously defines a particular foliation of spacetime into spacelike hypersurfaces.  (In an arbitrary tetrad frame gauge, this is not the case in general. For example, the tetrad lapse is, in general, smaller than the coordinate lapse due to the time dilation of the tetrad observer in the rest frame of $\Sigma_t$.)
It is important to note, though, that the representation of tetrad vector components through ADM variables does {\it not} mean the tetrad formulation is equivalent to the 3+1 ADM form. In particular, the tetrad formulation
can be rendered well-posed by an appropriate choice of gauge. By contrast, the 3+1 ADM formulation with {\it algebraic} gauge conditions (as commonly used in cosmology) cannot, which means the former can be implemented in numerical 
relativity but not the latter.

In terms of partial derivatives along the coordinate directions,
the directional derivatives along the {\it vierbein} take the simple form:
 \begin{equation}
D_0 = N^{-1} \Big( \partial_t - N^i\partial_i\Big)\quad {\rm and}\quad \quad D_a = E_a{}^i\partial_i
\,.
\end{equation}

\subsubsection{Coordinate gauge fixing}

The lapse function and the shift vector are gauge variables that together determine the particular foliation. In fixing the coordinate system, we have two goals: we want to choose a gauge that (i) renders the PDE system to yield a well-posed formulation and (ii) is well-adapted to the physical setting of contracting spacetimes. Most especially, the formulation should allow for studying spacetime contraction that lasts several hundreds of $e$-foldings.

Co-moving coordinates, {\it i.e.}, 
\begin{equation}
N^i \equiv 0\,,
\end{equation}
are a natural gauge choice, meaning that the spatial coordinates $x^i$ are constant along both the congruence and, due to the hypersurface orthogonal tetrad frame gauge, the foliation. For example, if the foliation is the same 
as the one used by observers,
the spatial coordinates do not introduce gauge artifacts.  

We fix the lapse $N$ by requiring that hypersurfaces of constant time $\Sigma_t$ are constant mean curvature (CMC) hypersurfaces. That is, the trace of the extrinsic curvature is spatially uniform on each $\Sigma_t$,
\begin{equation}
\Theta^{-1} \equiv  {\textstyle \frac13} K_a{}^a = {\rm const}.
\end{equation}
In the homogeneous and isotropic FRW limit, $\Theta$ is the Hubble radius $|H^{-1}|$. 
Choosing CMC slicing has several advantages:

First, it leads to a natural time coordinate choice
\begin{equation}
\label{time-choice}
e^t = {\textstyle \frac13} \Theta\,,
\end{equation}
with $t$ measuring the number of $e$-foldings of contraction of the Hubble radius. Note that, in the FRW limit, $t$ is related to the physical time coordinate $\tau$ through 
\begin{equation}
{\textstyle \frac13}e^{-t} = \frac{d\ln a(\tau)}{d\tau}\,.
\end{equation}

Second, it leads to a numerical scheme that is free of stiffness issues. The stiffness
problem arises becaue there are two dynamical variables, the Hubble radius and the scale factor, which decrease at exponentially different rates.  As noted above, 
in realistic scenarios of slow contraction, the Hubble radius decreases by a factor of $\sim 2^{50}$ during the same time 
that the scale factor shrinks by only a factor of two. By choosing CMC slicing that forces 
 $\Theta$ to be uniform and monotonic on slices of constant time $t$, we can eliminate the Hubble radius
 from the evolution equations by normalizing each dynamical variable by appropriate factors of $\Theta$, {\it i.e.}
\begin{eqnarray}
N &\rightarrow& {\cal N}\equiv N/\Theta,\\
\{K_{ab}, N_{ab}, E_a{}^i, \phi\} &\rightarrow&  \{\bar{K}_{ab}, \bar{N}_{ab}, \bar{E}_a{}^i, \bar{\phi} \} \,,\\
V &\rightarrow& \bar{V} \equiv V\times\Theta^2 \,,
\end{eqnarray}
where ${\cal N}$ is the Hubble-normalized lapse and bar denotes normalization by the mean curvature $\Theta^{-1}$ on constant time hypersurfaces.  

Third, the numerical simulation can run for any finite period without encountering singular behavior. 
With the time choice given in Eq.~\eqref{time-choice}, $t$ runs from small to large negative values.  The singular behavior occurs when $\Theta \rightarrow 0$, but this only occurs for $t\rightarrow -\infty$.   For any finite duration of the simulation, 
every curvature and each scalar field matter variable remains finite.  
 
\subsubsection{Evolution scheme}

Putting everything together, we obtain  the Einstein-scalar system~(\ref{Einstein-eq}-\ref{stress-energy}) in Hubble normalized, orthonormal tetrad form:
\begin{eqnarray}
\label{eq-E-ai-Hn}
\partial_t \bar{E}_a{}^i &=& - \Big({\cal N} - 1 \Big) \bar{E}_a{}^i - {\cal N} \,\bar{\Sigma}_a{}^b \bar{E}_b{}^i 
,\\
\label{eq-sigma-ab}
\partial _t \bar{\Sigma}_{ab} &=& - \Big( 3 {\cal N} - 1 \Big) \bar{\Sigma}_{ab}
- {\cal N} \Big( 2 \bar{n}_{\langle a}{}^c\, \bar{n}_{b \rangle c}
- \bar{n}^c{}_c \bar{n}_{\langle ab \rangle} 
- \bar{S}_{\langle a} \bar{S}_{b \rangle} \Big)
+ \bar{E}_{\langle a}{}^i\partial _i \Big(\bar{E}_{b \rangle}{}^i\partial _i {\cal N}\Big) 
\\
&-& {\cal N} \left( \bar{E}_{\langle a}{}^i \partial_i \bar{A}_{b \rangle}
-  \epsilon^{cd}{}_{(a} \Big( \bar{E}_c{}^i\partial_i \bar{n}_{b)d} - 2 \bar{A}_c \bar{n}_{b )d} \Big) 
  \right)
+  \epsilon^{c d}{}_{(a} \bar{n}_{b ) d} \bar{E}_c{}^i \partial_i{\cal N}
+ \bar{A}_{\langle a} \bar{E}_{b \rangle}{}^i \partial_i {\cal N} 
\nonumber
,\\
\label{eq-n-ab}
\partial _t \bar{n}_{ab} &=& - \Big({\cal N} - 1 \Big) \bar{n}_{ab} 
+ {\cal N} \Big( 2  \bar{n}_{(a}{}^c \bar{\Sigma}_{b)c}
-\epsilon^{cd}{}_{( a} \bar{E}_c{}^i \partial _i \bar{\Sigma} _{b ) d} \Big)   
- \epsilon^{cd}{}_{( a} \bar{\Sigma}_{b) d} \bar{E}_c{}^i \partial _i {\cal N} 
,\\
\label{eq-A-a}
\partial _t \bar{A}_a &=& - \Big( {\cal N} - 1 \Big)\bar{A}_a 
- {\cal N} \Big( \bar{\Sigma} _a{}^b \bar{A}_b - {\textstyle \frac12} \bar{E}_b{}^i\partial _i \bar{\Sigma} _a{}^b \Big)  
- \bar{E}_a{}^i \partial _i{\cal N} 
+ {\textstyle \frac12} \bar{\Sigma} _a{}^b \bar{E}_b{}^i \partial _i {\cal N} 
,\\
\label{eq-phi-Hn}
\partial_t \phi &=& {\cal N} \,\bar{W}
,\\
\label{eq-w-Hn}
\partial_t \bar{W} &=& - \Big(  3 {\cal N} -1 \Big) \bar{W} 
- {\cal N} \Big(\bar{V}_{,\phi}  + 2 \bar{A}^a \bar{S}_a -  \bar{E}_a{}^i \partial_i \bar{S}^a  \Big)
+ \bar{S}^a \bar{E}_a{}^i\partial _i {\cal N}
,\\
\label{eq-barS-Hn}
\partial_t \bar{S}_a &=& - \Big(  {\cal N} - 1 \Big) \bar{S}_a 
- {\cal N}\Big( \bar{\Sigma}_a{}^b \bar{S}_b - \bar{E}_a{}^i\partial _i \bar{W} \Big)
+  \bar{W} \bar{E}_a{}^i \partial_i {\cal N} 
,
\end{eqnarray}
where curved brackets denote symmetrization $X_{(ab)} \equiv {\textstyle \frac12}(X_{ab}+X_{ba})$ and angle brackets denote traceless symmetrization defined as $X_{\langle ab \rangle} \equiv X_{(ab)} - {\textstyle \frac13}X_c{}^c\delta_{ab}$. The geometric variables \begin{equation}
\bar{n}_{ab} \equiv \bar{N}_{(ab)}, \quad \bar{A}_b \equiv {\textstyle \frac12}\epsilon_b{}^{cd} \bar{N}_{cd}, 
\end{equation} 
are the symmetric and antisymmetric components, respectively of the Hubble-normalized, spatial curvature tensor $\bar{N}_{ab}$; $\bar{\Sigma}_{ab}$ is the trace-free extrinsic curvature tensor,
\begin{equation}
\bar{\Sigma}_{ab} \equiv \bar{K}_{ab} - 1.
\end{equation}
The scalar field matter variables 
\begin{equation}
\label{def-w-sa}
\bar{W}\equiv {\cal N}^{-1}\partial_t \phi ,\quad \bar{S}_a \equiv E_a{}^i\partial_i\phi,
\end{equation}
denote the Hubble-normalized field velocity and gradient of $\phi$, respectively. 

The evolution system~(\ref{eq-E-ai-Hn}-\ref{eq-barS-Hn}) is manifestly hyperbolic. The Hubble-normalized lapse, on the other hand, is subject to an elliptic equation,
\begin{eqnarray}
\label{Neqn}
&-& \bar{E}^a{}_i \partial^i \left(\bar{E}_a{}^j \partial _j {\cal N}\right) + 2 
\bar{A}^a \bar{E}_a{}^i\partial _i {\cal N} + {\cal N} \left(3 + \bar{\Sigma} _{a b}\bar{\Sigma}^{a b} + \bar{W}^2   - \bar{V} \right) = 
3 \,,
\end{eqnarray}
as a result of the CMC slicing condition. We are not aware of any rigorous proof of well-posedness for the particular tetrad formulation we use. See, however, Ref.~\cite{Andersson:2001kw} for a proof in a closely related coordinate based formulation which involves elliptic gauge conditions. 
Also, the fact that our code is stable and convergent is itself numerical
evidence that the underlying scheme is well-posed since otherwise one would expect to find instabilities or 
runaway behavior.

Finally, the geometric and scalar field matter tetrad variables satisfy the constraint equations 
\begin{eqnarray}
\label{constraintG}
&&3 + 2 \bar{E}_a{}^i \partial _a \bar{A}^a - 3 \bar{A}^a \bar{A}_a
- {\textstyle \frac12 } \bar{n}^{ab} \bar{n}_{ab}
+ {\textstyle \frac14 } ( \bar{n}^c{}_c)^2 
- {\textstyle \frac12 } \bar{\Sigma}^{ab} \bar{\Sigma}_{ab}
- {\textstyle \frac12 } \bar{W}^2 -  {\textstyle \frac12 } \bar{S}^a \bar{S}_a -   {\bar V} = 0
\,,\qquad\\
\label{constraintC}
&&\bar{E}_b{}^i \partial _i {\bar \Sigma} _a{}^b
 - 3 {\bar \Sigma} _a{}^b \bar{A}_b - \epsilon _a{}^{b c} \bar{n}_b{}^d \bar{\Sigma}_{cd} - {\bar W} {\bar S}_a = 0
\,,\\
\label{constraintJ}
&&\bar{E}_b{}^i \partial _i \bar{n}^b{}_a + \epsilon^{bc}{}_a \bar{E}_b{}^i\partial _i \bar{A}_c - 2 \bar{A}_b \bar{n}^b{}_a =0
\,,\\
\label{constraintS-phi}
&& {\bar S}_a - {\bar E}_a{}^i\partial _i\phi = 0
\,,\\
\label{constraintCOM}
&&\epsilon^{bc}{}_a
\Big( \bar{E}_b{}^j \partial_j \bar{E}_c{}^i - \bar{A}_b \bar{E}_c{}^i \Big) - \bar{n}_a{}^d \bar{E}_d{}^i = 0.
\end{eqnarray}
As detailed in the following, we will utilize the constraint equations to specify the initial conditions as well as to check for numerical convergence.

\section{Initial conditions}
\label{sec_initial}

The Einstein-scalar field equations must be supplemented by initial conditions that satisfy the Hamiltonian and momentum constraints. 
By construction, our scheme allows for the variation of all freely specifiable geometric and scalar-matter field variables, $\{\bar{n}_{ab}, \bar{A}_b, \bar{\Sigma}_{ab}, \bar{E}_a{}^i  \}$ and $\{\phi, \bar{W}\}$, and therefore enables us to study slow contraction under a wide range of initial conditions, in particular those that lie far outside the perturbative regime of FRW spacetimes. 

As described in Ref.~\cite{Ijjas:2020dws}, to specify the geometric variables $\{\bar{n}_{ab}, \bar{A}_b, \bar{\Sigma}_{ab}, \bar{E}_a{}^i  \}$ at some initial time $t_0$, we shall employ the so-called York method \cite{York:1971hw} as commonly used in numerical general relativity. In particular, we choose the spatial metric of the $t_0$-hypersurface to be conformally-flat,
\begin{equation}
\tensor{g}{_i_j}(t_0, \vec{x}) = \psi^4(t_0, \vec{x}) \tensor{\delta}{_i_j};
\end{equation}
where the conformal factor $\psi$ is not a free function but obeys an elliptic equation by the Hamiltonian constraint~\eqref{conformal}, as described below. 
Together with the constant mean curvature $\Theta^{-1}_0$ of the ${t_0}$-hypersurface, which we freely specify, this choice for the spatial metric fixes the coordinate components of the spatial triad
\begin{equation}
\tensor{\bar{E}}{_a^i} = \psi^{-2} \Theta_0^{-1} \tensor{\delta}{_a^i}\,;
\end{equation}
as well as  all {\it intrinsic} curvature variables 
\begin{equation}
\tensor{\bar{n}}{_a_b} (t_0, \vec{x})= 0\,, \quad {\rm and} \quad \tensor{\bar{A}}{_b}( t_0, \vec{x}) = -2 \psi^{-1}(t_0, \vec{x}) \tensor{\bar{E}}{_b^i}(t_0, \vec{x})  \tensor{\partial}{_i} \psi(t_0, \vec{x})\,.
\end{equation}
Note, though, that conformal flatness does not mean zero intrinsic curvature for the $t_0$-hypersurface. Rather, since the anti-symmetric part of the intrinsic curvature does not transform trivially under conformal rescaling, $\bar{A}_b\neq 0$ in general. 

Furthermore, the momentum constraint~\eqref{constraintC} reduces to the simple expression:
\begin{equation}
\label{init-momentum-c}
\tensor{\bar{E}}{^a_i} (t_0, \vec{x})  \partial^i \tensor{Z}{_a_b}(t_0, \vec{x}) = Q(t_0, \vec{x})  \tensor{\bar{E}}{_b^i}(t_0, \vec{x})  \partial_i \phi(t_0, \vec{x}) \,,
\end{equation}
relating the trace-free part of the conformally-rescaled {\it extrinsic} curvature (or shear) tensor,
\begin{equation}
\tensor{Z}{_a_b}(t_0, \vec{x}) = \psi^6(t_0, \vec{x}) \tensor{\bar{\Sigma}}{_a_b}(t_0, \vec{x})\,,
\end{equation}
with the scalar field matter variables, the field distribution $\phi(t_0, \vec{x})$ and its conformally rescaled Hubble-normalized velocity distribution on the $t_0$-hypersurface,
\begin{equation}
\label{Def-Q}
Q(t_0, \vec{x})  = \psi^6(t_0, \vec{x}) \bar{W}(t_0, \vec{x}).
\end{equation}
Eq.~\eqref{init-momentum-c} makes it possible to freely specify both matter variables $\{\phi, Q \}$ as well as the vacuum contribution $\tensor*{Z}{_a_b^0}$ of the Hubble-normalized extrinsic curvature tensor, which is independent of the matter source,
\begin{equation}
\label{init-momentum-c0}
 \partial^a \tensor*{Z}{_a_b^0} = 0 \,.
\end{equation}
The momentum constraint~\eqref{init-momentum-c} yields the rest of the initial shear contribution $\tensor{Z}{_a_b}-\tensor*{Z}{_a_b^0}$ . 

The logic behind choosing the initial data for $\phi, Q$ and $\tensor*{Z}{_a_b^0}$ is detailed in Refs.~\cite{Garfinkle:2008ei,Ijjas:2020dws}. Here, we go beyond previous work in that we allow deviations from homogeneity in {\it two} spatial directions $x$ and $y$. We define the initial vacuum shear contribution to be given by
\begin{equation}
\tensor*{Z}{_a_b^0} = 
\begin{pmatrix}
b_2 + c_2 \cos{y} & {\;} & \xi & {\;} &  \kappa_1+ c_1 \cos{y}  \\
&&&&\\
\xi & \; &  b_1+  a_1 \cos{x} & {\;} &  \kappa_2 +a_2 \cos{x}  \\
&&&&\\
\kappa_1+ c_1 \cos{y} 
&  {\;} &  \kappa_2+ a_2 \cos{x}  &  {\;} & -b_1 - b_2 - a_1 \cos{x} - c_2 \cos{y}
\end{pmatrix}
\label{ZZ}
\end{equation}
where $a_1, a_2, b_1, b_2, c_1, c_2, \kappa_1, \kappa_2$ are constants; and we fix the scalar field variables as follows: 
\begin{align}
\begin{split}
\label{QQ}
Q &= \Theta \Big(q_{x} \cos{(m_x x + d_x)} + q_{y} \cos{(m_y y + d_y)} + Q_0 \Big) \\
\phi &= f_x \cos{(n_x x + h_x)} + f_y \cos{(n_y y + h_y)} + \phi_0\,,
\end{split}
\end{align}
where $Q_0, \phi_0, q_x, q_y, f_x, f_y, m_x, m_y, n_x, n_y, d_x, d_y, h_x, h_y$ are constant and denote the mean value, the amplitude, the mode number and the phase of the initial  velocity and field distribution, respectively. The choice of cosine reflects the fact that, for the numerical simulation, we choose periodic boundary conditions $0\leq x,y\leq2\pi$ with $0$ and $2\pi$ identified.

Finally, putting everything together, the conformal factor $\psi(t_0, \vec{x})$ is numerically computed from  the Hamiltonian constraint~\eqref{constraintG}, which yields an elliptic equation for $\psi$:
\begin{eqnarray} 
\label{conformal}
\partial^i \partial _i \psi &=& {\textstyle \frac14 } \left( 3 \Theta^{-2} - V  \right) \psi^5 - {\textstyle \frac18 } \left( \partial^i \phi \partial _i \phi \right)\psi
- {\textstyle \frac18 } \left( Q^2  + Z^{ab} Z_{ab} \right) \Theta^2 \psi^{-7}.
\end{eqnarray}

\section{Numerical analysis}
\label{sec_numanalysis}

In Ref.~\cite{Ijjas:2020dws}, the   evolution scheme described in Sec.~\ref{sec_scheme} was applied in
 a numerical relativity code that accepts initial conditions with non-perturbative deviations from homogeneity and isotropy along {\it a single spatial direction}.  
 This section presents simulations using an improved code that accepts initial conditions with deviations from homogeneity and isotropy along {\it two independent spatial directions} as described in Sec.~\ref{sec_initial}.  We will refer to the earlier and new simulations as having  {\it one- and two-dimensional initial conditions}, respectively.   Note  that 
 both codes evolve the full $(3+1)$-dimensional Einstein-scalar field system of equations; it is only the dimensionality of the initial conditions that differ. 
 
We will begin by constructing three representative simulations with two-dimensional initial conditions to test whether the smoothing due to slow contraction is qualitatively  similar to the extraordinarily robust smoothing effect found 
using our earlier code with one-dimensional initial conditions.  We will then use the numerical code with two-dimensional initial conditions in this and forthcoming papers to perform an extensive series of systematic studies of:
\begin{itemize} 
\item the role of ultralocality in the smoothing process (in Sec.~\ref{sec_ultralocal}); 
\item the effects of mode coupling on the robustness to initial conditions and the rapidity of the smoothing process when there are deviations from homogeneity in more than one direction (in Ref.~\cite{MCpaper}); 
\item and, the `spike' phenomenon, {\it i.e.}, rapid ``small scale spatial structure'' variations  first reported in Ref.~\cite{Berger:1993ff} for numerical general relativity simulations of vacuum spacetimes and more recently observed in some of our simulations with Einstein gravity coupled to a  scalar field (in Ref.~\cite{Spikepaper}).
\end{itemize} 

Aside from extending the numerical code used in Ref.~\cite{Ijjas:2020dws} to enable initial conditions with deviations in two independent spatial directions, the only notable change in the new code is the use of the multigrid V-cycle method to solve more efficiently the elliptic equation~\eqref{conformal} for the lapse.  The multigrid V-cycle method (along with other methods) was first tested on the original code (with one-dimensional initial conditions) to verify that all methods agreed.  In addition, as described in Appendix~A, we have introduced more cross-checks for code  convergence.  

A key result of the numerical studies in Ref.~\cite{Ijjas:2020dws} based on one-dimensional initial conditions was that slow contraction with equation of state  $\varepsilon\gtrsim 13$ is powerfully robust (smoothing and flattening for initial conditions with large non-perturbative deviations from FRW)
and rapid (accomplishing the feat by the time the Hubble radius shrinks by only a few $e$-folds).  The result, based on the outcomes of many hundreds of simulations, was summarized in a series of phase diagrams that depend on $Q_0$, the spatially averaged initial  field-velocity  in Eq.~(\ref{QQ}), and $\varepsilon$, the equation of state parameter.  

As discussed in Ref.~\cite{Cook:2020oaj}, the phase behavior is especially sensitive to  $Q_0$.  Positive values of $Q_0$ correspond to the average initial field velocity being directed down the steep exponential potential that drives the slow contraction, the condition that naturally occurs in bouncing and cyclic models, and, hence, the case of practical interest for cosmology.  However, to fully understand the range over which  slow contraction is an effective smoothing mechanism and to study other effects of interest in general relativity, we also consider here and in subsequent studies cases where the average initial velocity is nearly at rest or headed `wrongway' ($Q_0 \lesssim 0$).

\begin{figure}[t]
\center{\includegraphics[width=4.25in,angle=-0]{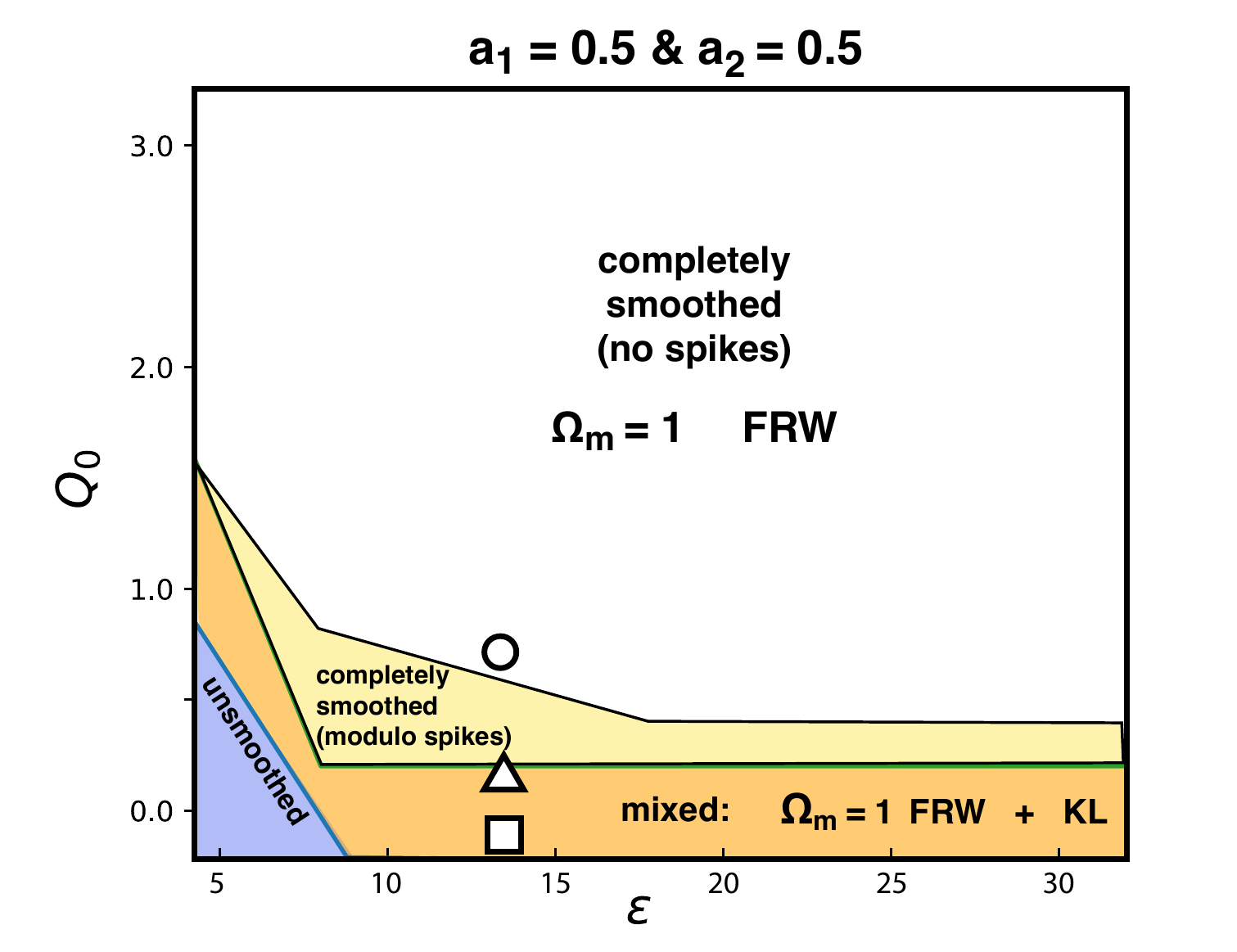}} 
\caption{Phase diagram adapted from Fig.~10 in  Ref.~\cite{Ijjas:2020dws}  showing the final states reached beginning
from  a combination of non-perturbative inhomogeneities in both the field velocity and shear degrees of freedom.  
The entire region relevant
to cyclic and bouncing cosmology models --  $\varepsilon>13$ and $Q_0 \ge 0$ -- converges completely or to an
exponential degree (as measured by proper volume) to the desired smooth, anisotropic and
flat dynamical attractor solution with spikey behavior in the regimes indicated. 
Superposed on the diagram are markers indicating
three cases studied in this paper using the new code with initial spatial variations along two independent spatial directions: $Q_0=0.7$, complete smoothing without spikes (circle); $Q_0=0.1$, complete smoothing but with some spiking before smoothing (triangle); and, $Q_0=-0.1$, a mixed outcome of smoothed and exponentially small (as measured by proper volume) unsmoothed regions with spiking.}
 \label{Figure1}
\end{figure}

Fig.~10 in  Ref.~\cite{Ijjas:2020dws} contains a phase diagram showing the outcome for initial states with
 large non-perturbative deviations from homogeneity and isotropy
 (along one dimension) in both the scalar field and shear degrees of freedom.
 We reproduce a  version of this diagram in Fig.~\ref{Figure1}. 

Using the new code that allows initial deviations from homogeneity and isotropy along two independent spatial directions, we have run representative simulations to test whether there are substantial changes in the phase diagram.  In this section, for example, we present three cases indicated by the three symbols in Fig.~\ref{Figure1}.   All three correspond to a scalar field potential $V(\phi) = V_0  \exp(- \phi/M)$, as introduced above in Eq.~\eqref{V-def}, with $V_0=-0.1$ and $\varepsilon= 1/(2M^2) =13$ (assuming reduced Planck units) and to 
 $a_1=0.5$, $a_2=0.5$, $b_1=-0.15$, $b_2=1.8$, $\xi=0.01$, $q_x=0.51 $, and $d_x=0$, as defined in Eqs.~(\ref{ZZ}) and~(\ref{QQ}).   These parameters, which describe the potential and the initial spatial variation along the $x$ direction, were intentionally chosen to be  the same as the example used in Fig.~\ref{Figure1} in Ref.~\cite{Ijjas:2020dws}.  The one change along the $x$ direction compared to Ref.~\cite{Ijjas:2020dws} is that a higher mode  spatial variation, $m_x=2$ rather than $m_x=1$, is chosen  to test whether this makes a qualitative difference to the phase diagram.  In the new code, the three examples also have   non-trivial initial deviations along the $y$ direction corresponding to $c_1=0.5$, $c_2=0.5$, and  $\kappa_1 =\kappa_2=0.01$ and $m_y=3$ in Eqs.~(\ref{ZZ}) and~(\ref{QQ}).  Just as in the $x$ direction, these parameters correspond to highly non-perturbative deviations away from the slow contraction attractor solution reached in the smoothed regions. 
 (Both the old and new code also have $\phi(\vec{x},t=0)=0$; the outcomes are relatively insensitive to the initial field values.)  

\begin{figure}[t]
\begin{center}
\center{\includegraphics[width=6.0in,angle=-0]{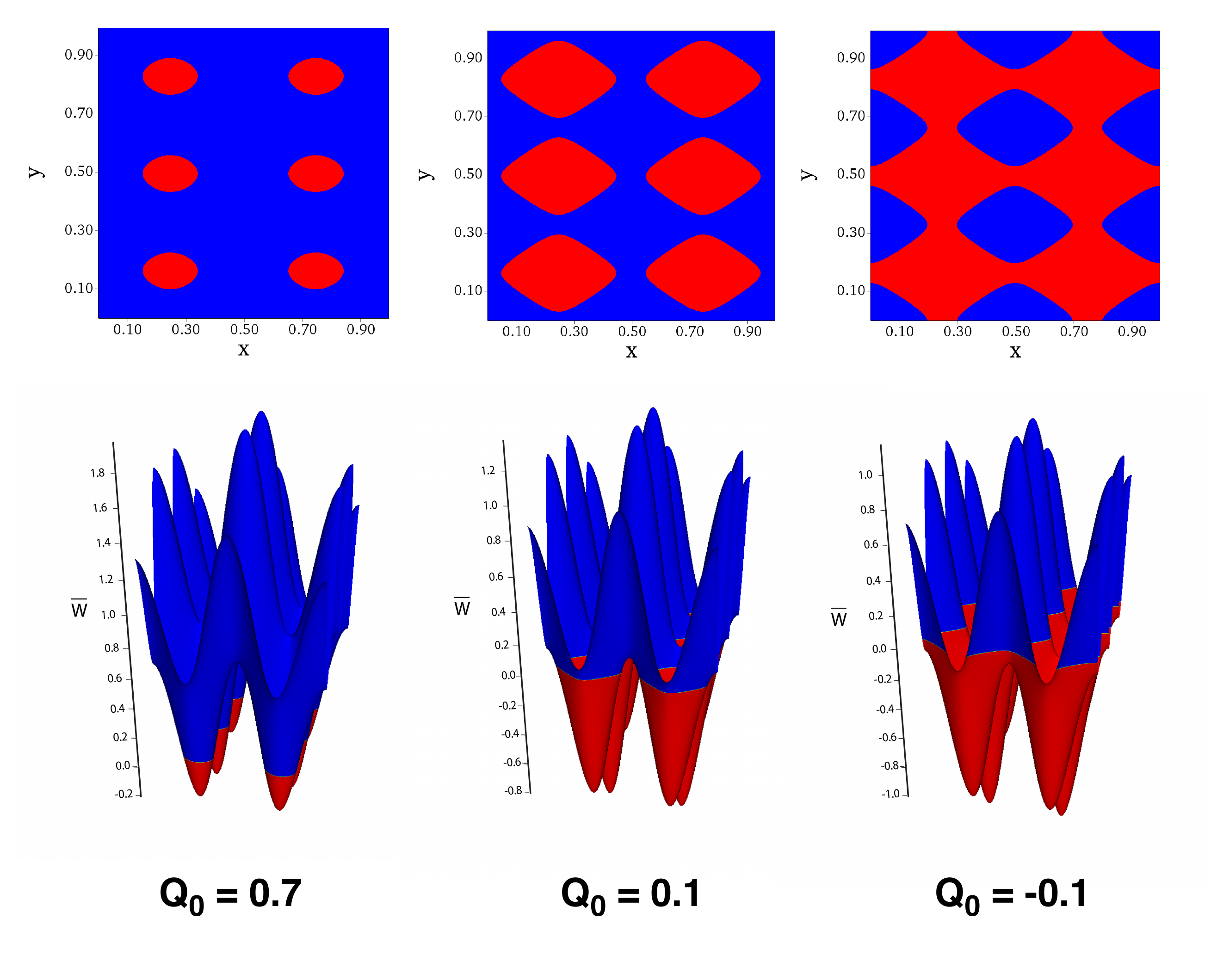}}
\caption{The initial $\bar{W}$-configuration for the three test cases with average initial field-velocity $Q_0 = \{ 0.7, \, 0.1,\, -0.1\}$ that 
correspond to the circle, triangle and square, respectively, in Fig.~\ref{Figure1}, where the relation between $\bar{W}$ and $Q$ is given in Eq.~\eqref{Def-Q}. The first row shows a top view that only distinguishes downhill (blue) from uphill (red) initial field-velocity; there is no spatial variation along the $z$ direction.   The second row shows $\bar{W}(\vec{x},t=0)$ as a function of $\vec{x}= (x,\, y)$ where the division between blue and red corresponds to $\bar{W}(\vec{x},t=0)=0$. }
 \label{Figure2}
 \end{center}
\end{figure}

The only difference among the three cases presented in this section is the average initial field velocity parameter, $Q_0$, as shown in Fig.~\ref{Figure2}. This leads to the three different plots of  $\bar{W}(\vec{x},t=0)$, where blue represents initial velocities aimed down the potential (rightway) and red correspond to initial velocities aimed up the potential (wrongway).  As $Q_0$ is decreased from 0.7 (circle), to 0.1 (triangle) to -0.1(square), the fraction of space in which the field velocity is aimed in the wrong direction increases.   Because different modes $m_x=2$ and $m_y=3$ have been chosen along the two spatial directions, the initial field configuration has six regions with  substantial wrongway initial conditions.  Note, though, that the figures do not show the initial shear, which does not have the same symmetry as the $Q$-field
shown in Fig.~\ref{Figure2}.

Let us emphasize that, in bouncing and cyclic models, a necessary condition to begin any type of contraction is that the scalar field velocity is aimed down a negative potential ({\it i.e.}, rightway) everywhere. In our study here, as in Ref.~\cite{Ijjas:2020dws}, we are intentionally exploring `unnatural' conditions in which the initial scalar
field velocity is aimed uphill (wrongway) in some regions of spacetime.
Our aim in these cases it to examine if the power of slow contraction is substantially different. 
Furthermore, we want to examine phenomena that can occur when smoothing is not complete. 

\begin{figure}[t]
\begin{center}
\center{\includegraphics[width=6.25in,angle=-0]{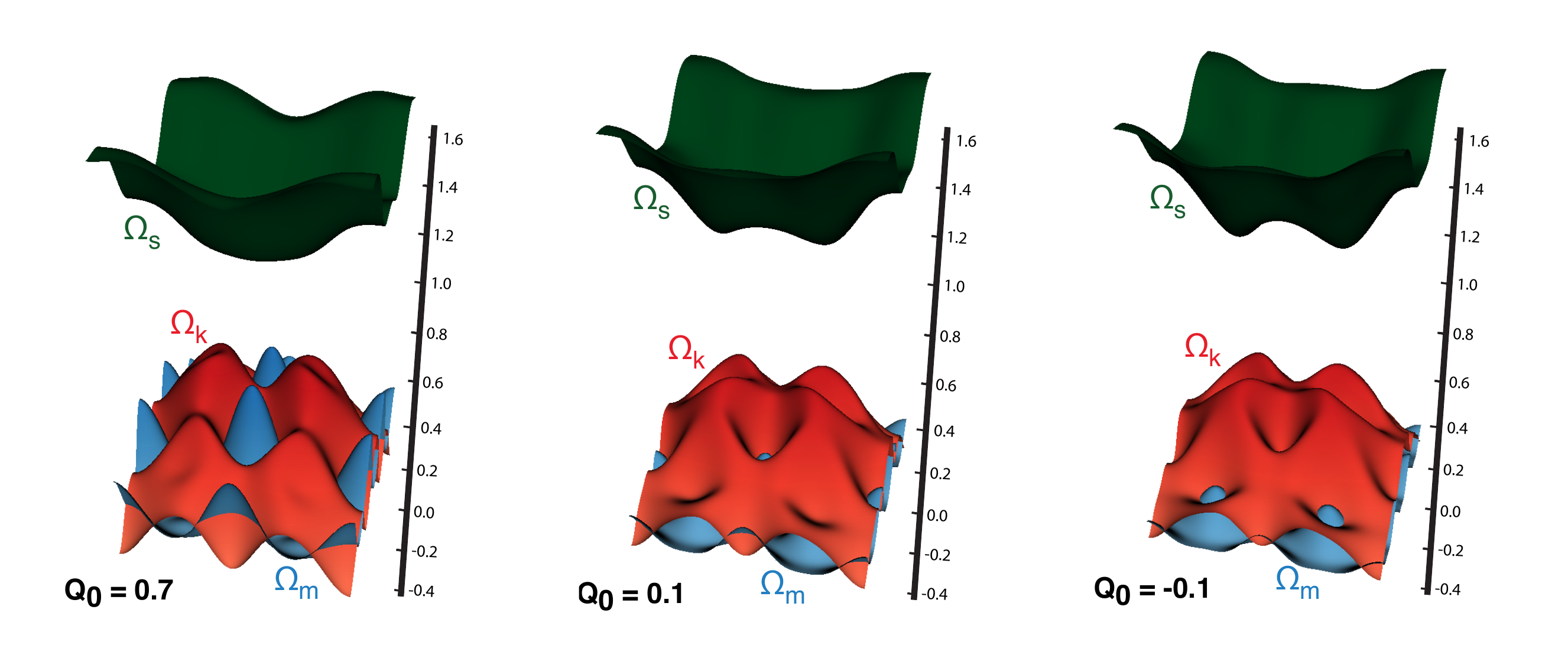}}
\caption{The initial  spatial distributions of the three energy density components, $m=$~matter (blue), $k=$~spatial curvature (red) and $s=$~shear (green), for the three case studies: $Q_0 = \{ 0.7, \, 0.1,\, -0.1\}$. Note that at $t=0$ the  shear  (green) dominates and matter and spatial curvature contributions are subdominant. }
 \label{Figure3}
 \end{center}
\end{figure}

Figure~\ref{Figure3} shows the initial inhomogeneous and anisotropic spatial variations of the $\Omega_i$, the ratios of the densities of the three energy components divided by $3/\Theta^2$ ($m=$~matter, $k=$~spatial curvature and $s=$~shear),
{\it i.e.},
\begin{eqnarray}
\Omega_m& \equiv & {\textstyle\frac16} \bar{W}^2 + {\textstyle\frac16} \bar{S}^a \bar{S}_a + {\textstyle\frac13} \bar{V} ,
\\
\Omega_k  & \equiv & - {\textstyle\frac23} \bar{E}_a{}^i\partial_{i} \bar{A}^a +  \bar{A}^a \bar{A}_a + {\textstyle\frac16} \bar{n}^{ab} \bar{n}_{ab} - {\textstyle\frac{1}{12}}(\bar{n}^c{}_c)^2,
\\
\Omega_s & \equiv & {\textstyle\frac16} \bar{\Sigma}^{ab} \bar{\Sigma}_{ab}  .
\end{eqnarray}
(Matter in this paper always refers to the scalar field energy density.)  The sum of the $\Omega_i$ is constrained to equal unity on constant mean curvature hypersurfaces.

\begin{figure}[t]
\begin{center}
\includegraphics[width=1.0\linewidth]{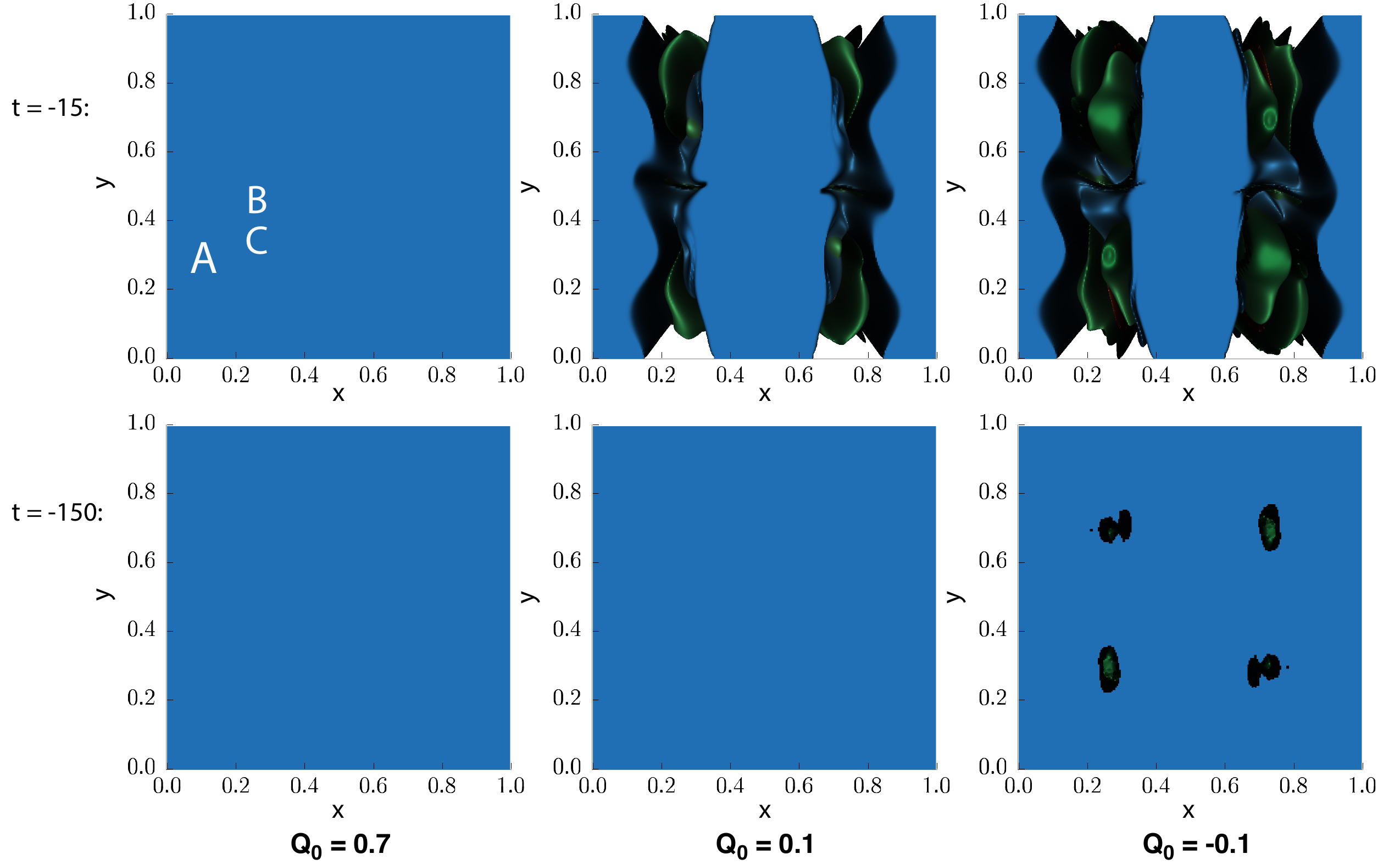}
\caption{The spatial distributions of the three energy density components ($m=$~matter, $k=$~spatial curvature and $s=$~shear) for the three models $Q_0 = \{ 0.7, \, 0.1,\, -0.1\}$ after $t_{\rm spike}=-15$ (top row) and $t_{\rm end}=-150$.  Blue indicates smoothed regions of spacetime dominated by the scalar field energy density ($\Omega_m=1$); all other shades correspond to unsmoothed regions with an inhomogeneous mixture of the three components. The behavior three points marked $A$, $B$ and $C$ in the top-left panel will be 
discussed in Sec.~\ref{sec_ultralocal}
}
 \label{Figure4}
\end{center}
\end{figure}

Figure~\ref{Figure4} shows the  $\Omega_i$ at an early time $t_{\rm spike} = -15$ and at the end of the slow 
contraction phase, $t_{\rm end}=-150$, respectively.  Recall that $t$ as defined in Eq.~\eqref{time-choice} measures the number of $e$-foldings of contraction of the Hubble radius $\Theta$. The times correspond to 15 and 150 e-folds of contraction of the Hubble radius.  Since $a(t) \propto \Theta^{1/\varepsilon}$, there is negligible contraction of the scale factor in either case, as is characteristic of slow contraction.  The value $t_{\rm end}=-150$ was chosen because the total duration of 
slow contraction in bouncing and cyclic models is in the range  $\Delta t= 100-150$, depending on details of the model.
For the $Q_0 =0.7$ and $Q_0 =0.1$ cases (for which the average initial field-velocity is downhill directed), the final outcome is smooth with $\Omega_m=1$ everywhere.  For the $Q_0=-0.1$ case, the final outcome is a mixture of smoothed and unsmoothed regions.  Note that the four unsmoothed regions do not correspond to the (red) regions with uphill initial field-velocity in Fig.~\ref{Figure2}; this is due to the initial shear that breaks the symmetry of the field-velocity distribution, as noted above, a sign that we are observing inhomogeneities and anisotropies evolving in two independent spatial directions.

The impressive result is that the three behaviors found using the code agree well with what was found using the previous code Ref.~\cite{Ijjas:2020dws}  despite the fact that the initial conditions are here two-dimensional.  This is a first sign that there are no significant differences when initial deviations from FRW along two independent directions are introduced, even though one might have imagined mode coupling and other non-linear multi-dimensional effects could come into play.  Similarly, the fact that no spiking was observed in the $Q_0=0.7$ case and that the smoothing is complete (as observed in Fig.~\ref{Figure4}) before $t_{\rm spike}=-15$ (that is, after less than ten percent 
of the total smoothing period) is indicative that the robustness and rapidity of slow contraction
 remains over most of phase space (the white region in Fig.~\ref{Figure1})
 when deviations in two independent directions are included.  

We are also interested in extreme cases of initial conditions where more complex behavior occurs.  Both the $Q_0 =0.1$ and $Q_0 =-0.1$ show small but detectable deviations from convergence in some small isolated regions, as we will
detail below,  that are maximum around $t_{\rm spike} = -15$ (see Appendix A). This signifies spiking behavior, which is why we show snapshots at this time as well. Note that, with substantial wrongway  initial field velocity, neither case is smooth at $t_{\rm spike} = -15$, though complete 
smoothing appears to be achieved by $t_{\rm {end}}=-150$ $e$-folds of slow contraction in the $Q=0.1$ case.

\section{Ultralocality}
\label{sec_ultralocal}

In this section, we use the three cases described in Sec.~\ref{sec_numanalysis} to examine the `detailed process' by which slow contraction `smooths' beginning from highly non-perturbative deviations from a flat FRW spacetime.  

By `smooths,' we mean converging to a homogeneous and isotropic geometry well-described by a flat FRW metric with $\Omega_m =1$.  We use the term `smoothing' by itself to describe convergence to flat FRW  by the time $t_{\rm end}$ in either a local region or set of regions in a simulation; `complete smoothing'  refers to convergence for all spacetimes points 
by $t_{\rm end}$.

By `detailed process,' we mean to reveal how the smoothing in any particular local region depends
on the local initial conditions.  The conclusions are based
on examining the behavior of four indicators:  
\begin{itemize}
\item the symmetric component of the Hubble-normalized spatial curvature $\bar{n}_{ab}(t)$ 
Eq.~(\ref{eq-n-ab}); 
\item the Hubble-normalized trace-free extrinsic curvature tensor $\bar{\Sigma}_{ab}(t)$ in Eq.~(\ref{eq-sigma-ab}); 
\item the Hubble-normalized scalar field velocity $\bar{W}(t)$ in Eq.~(\ref{eq-w-Hn});
and, 
\item the Hubble-normalized lapse ${\cal N}(t)$ in Eq.~(\ref{Neqn}).  
\end{itemize}
Their time-variation depends on `velocity' terms that do not involve spatial derivatives and `gradient' terms that explicitly involve spatial derivatives \cite{Belinsky:1970ew}.  
 Smoothing to the flat FRW attractor solution
 in any local region corresponds to having the time-variation of $\bar{n}_{ab}$,
 $\bar{\Sigma}_{ab}$, and $\bar{W}$ as well as $\bar{n}_{ab}$,
 $\bar{\Sigma}_{ab}$ approach zero to within the resolution limit of the simulation (about $10^{-6}$)
 and ${\cal N}$ approaches $1/\varepsilon$ \cite{Lim:2009dg,Ijjas:2020dws}.

As examples, we use the three representative 
simulations 
described in Sec.~\ref{sec_numanalysis}.
Each simulation
begins with the same initial conditions except for the average initial scalar field velocity $Q_0$:
\begin{itemize}
\item $Q_0=+0.7$:  representative of most of phase space in Fig.~\ref{Figure1}, completely smooths without any sign of spiking behavior;
\item $Q_0=+0.1$: exhibits spiking in some localized regions but appears to completely smooth by $t_{\rm end}$; and,
\item $Q_0=-0.1$: exhibits spiking  and does not smooth in some localized regions by $t_{\rm end}$.
\end{itemize}
Figure~\ref{Figure2} shows the top view looking down the $z$ direction; the initial conditions of our (3+1)D simulations vary spatially in the $x-y$ plane  and are uniform along any line along the $z$ direction.   For each of the three simulations, we will focus on the 
three representative lines indicated in the upper-left-most top-view of Fig.~\ref{Figure4}, labeled:
\begin{itemize}
\item $A=(0.1, 0.28)$: a point that appears to go directly to a smooth phase in all three cases above;
\item $B=(0.25, 0.45)$: a point that appears to go directly to a smooth phase in the first case above ($Q_0=0.7$), but only smooths in the last two cases ($Q_0=\pm 0.1$)  following a more complex evolution; and
 \item $C=(0.25, 0.33$): a point that appears to go directly to a smooth phase for $Q_0=0.7$; only smooths for $Q_0=0.1$ following a more complicated evolution; does not smooth at all for $Q_0=-0.1$.
\end{itemize}

 \begin{figure}
\begin{center}
\includegraphics[width=0.3\linewidth]{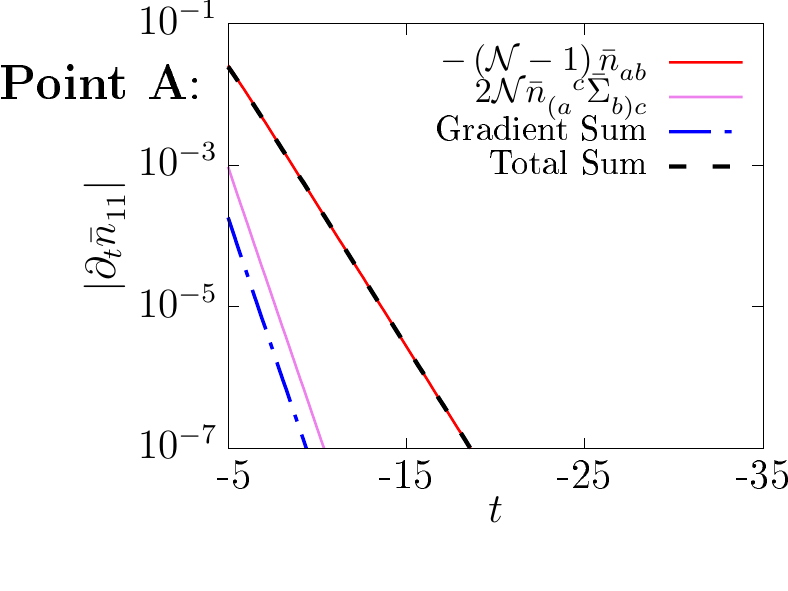}
\includegraphics[width=0.3\linewidth]{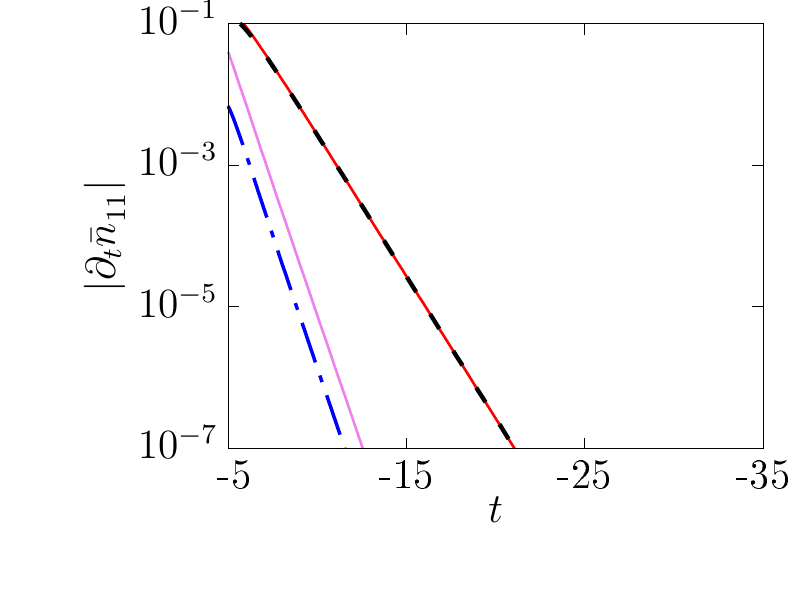}
\includegraphics[width=0.3\linewidth]{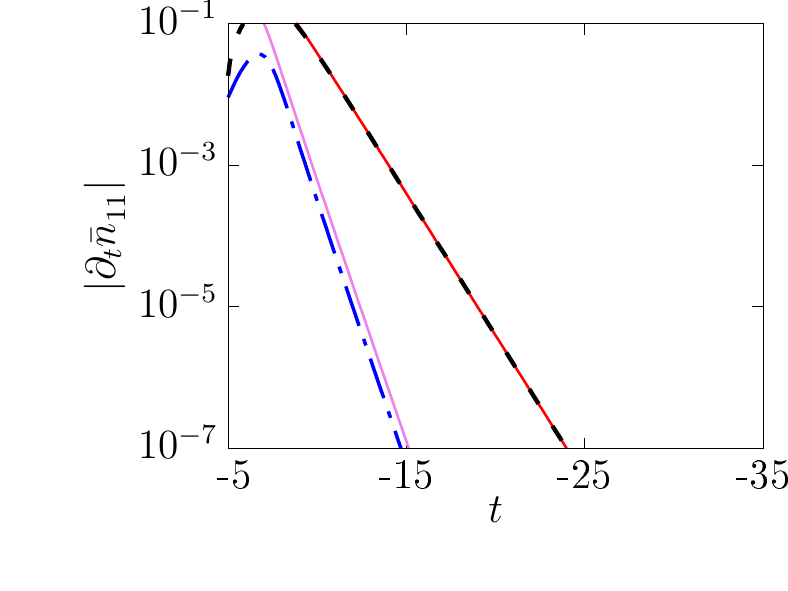} \\
\includegraphics[width=0.3\linewidth]{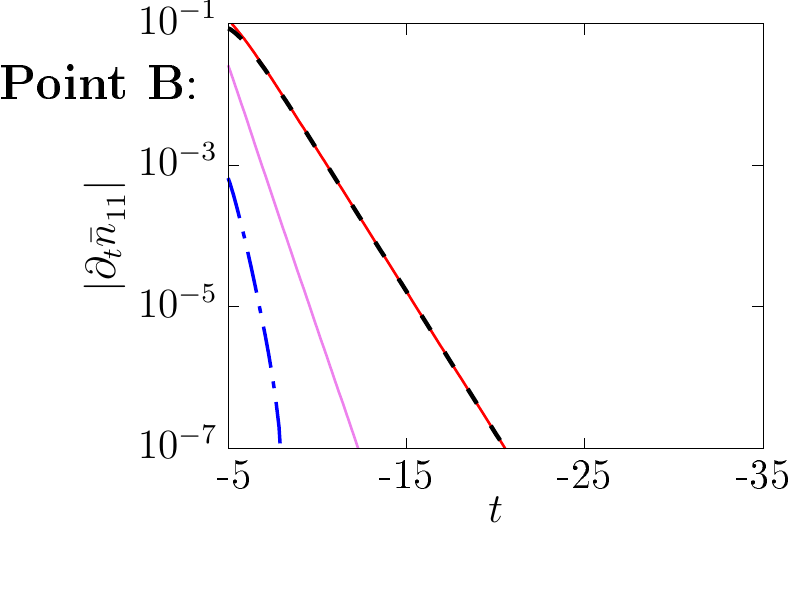}
\includegraphics[width=0.3\linewidth]{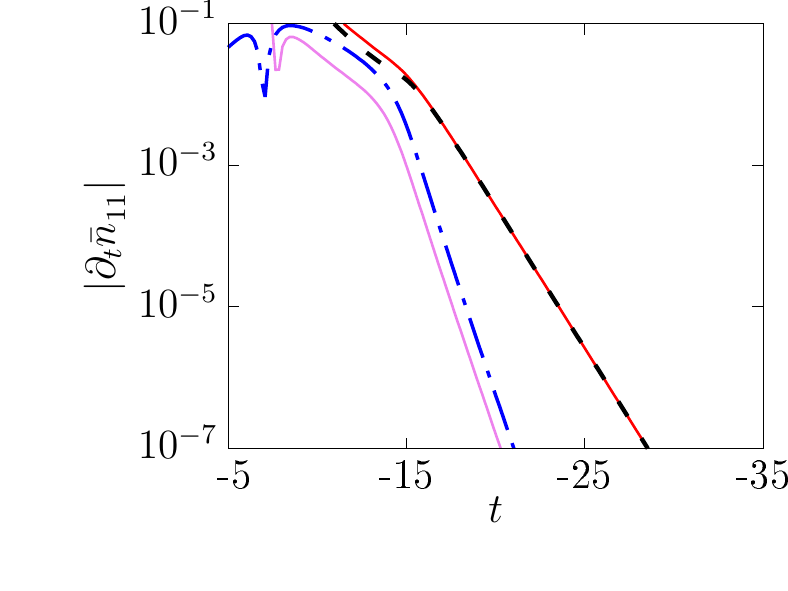}
\includegraphics[width=0.3\linewidth]{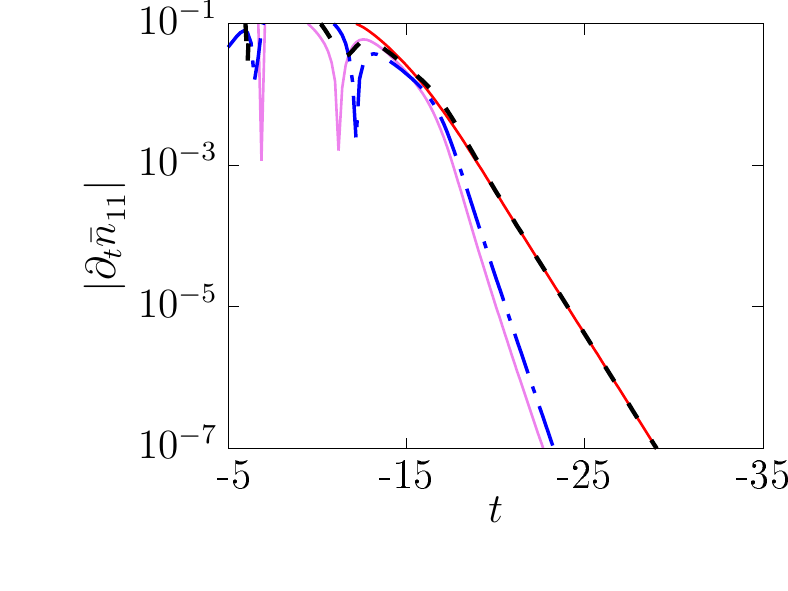} \\
\includegraphics[width=0.3\linewidth]{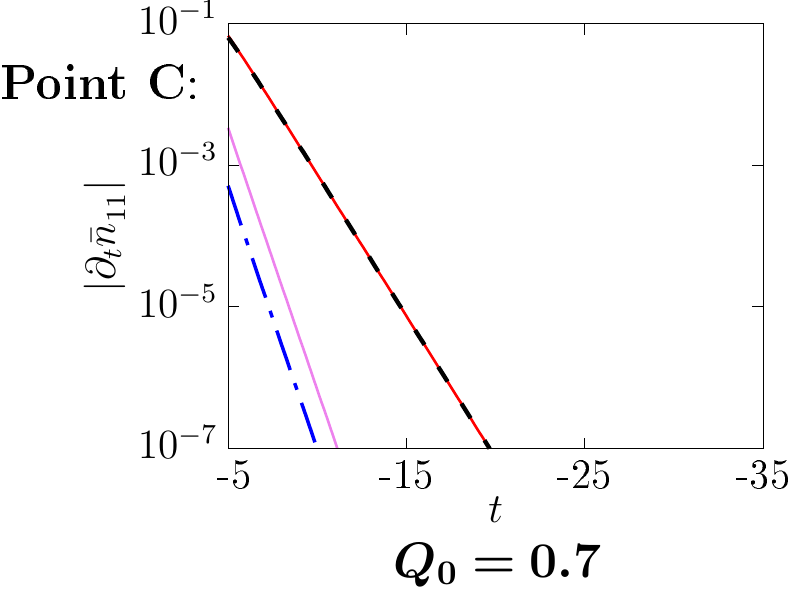}
\includegraphics[width=0.3\linewidth]{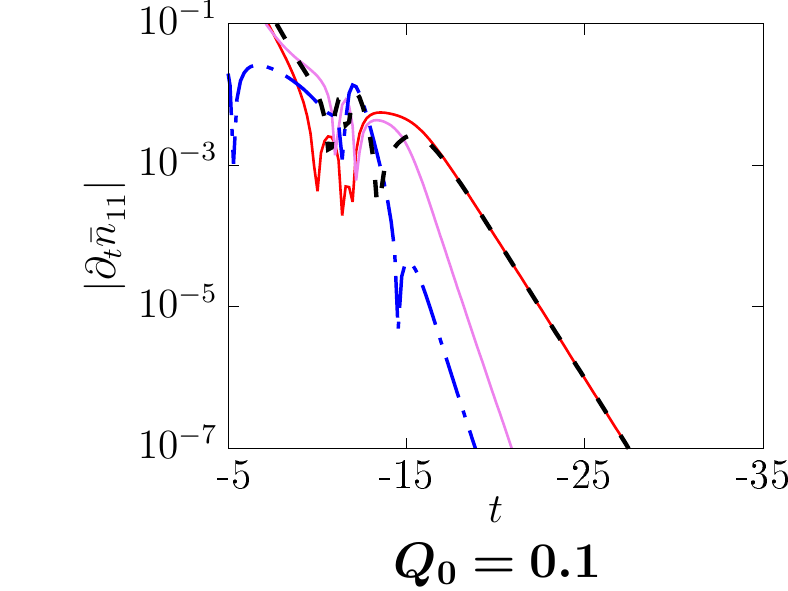}
\includegraphics[width=0.3\linewidth]{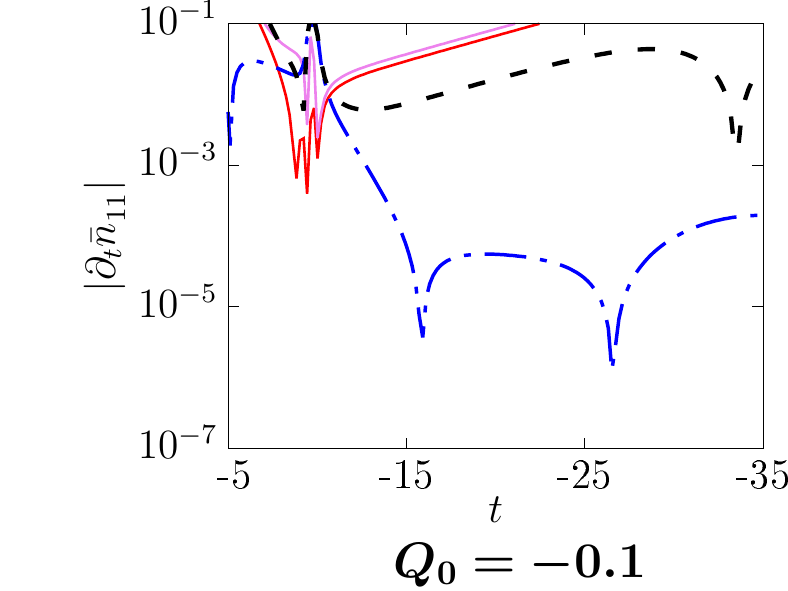} \\
\caption{
Evolution of $|\partial_t \tensor {\bar{n}}{_1_1}|$: columns correspond to $Q_0=0.7$(left); $Q_0=0.1$ (middle), and $Q_0=-0.1$ (right). Rows correspond to points  $A$ (top), $B$ (middle), and $C$ (bottom), as described in the text.
The different curves correspond to the magnitudes of different contributions, as indicated in the legend, 
to the $(11)$-component of Eq.~\eqref{eq-n-ab}.
}
 \label{Figure5}
\end{center}
\end{figure}

\begin{figure}
\begin{center}
\includegraphics[width=0.3\linewidth]{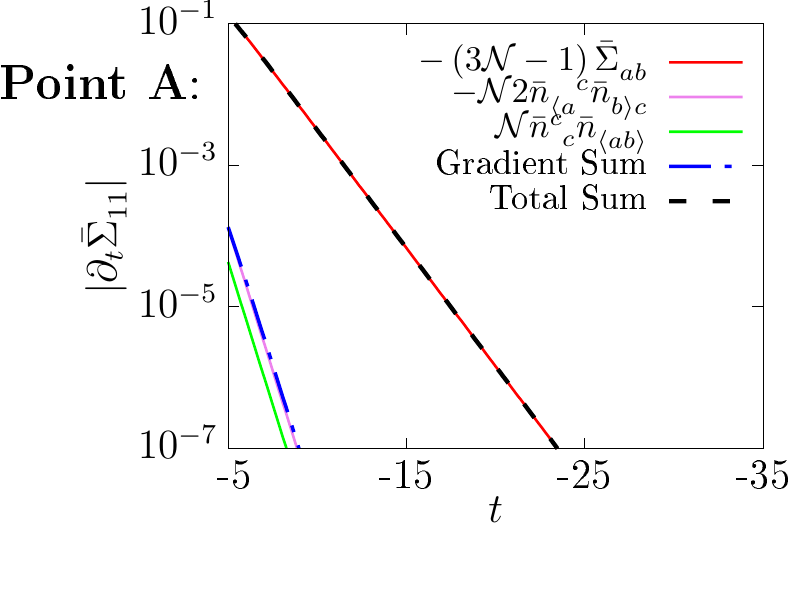}
\includegraphics[width=0.3\linewidth]{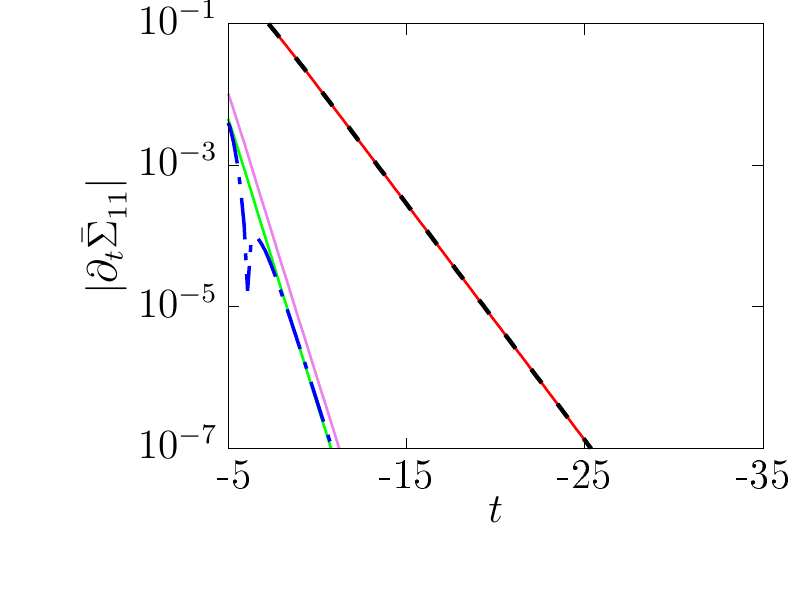}
\includegraphics[width=0.3\linewidth]{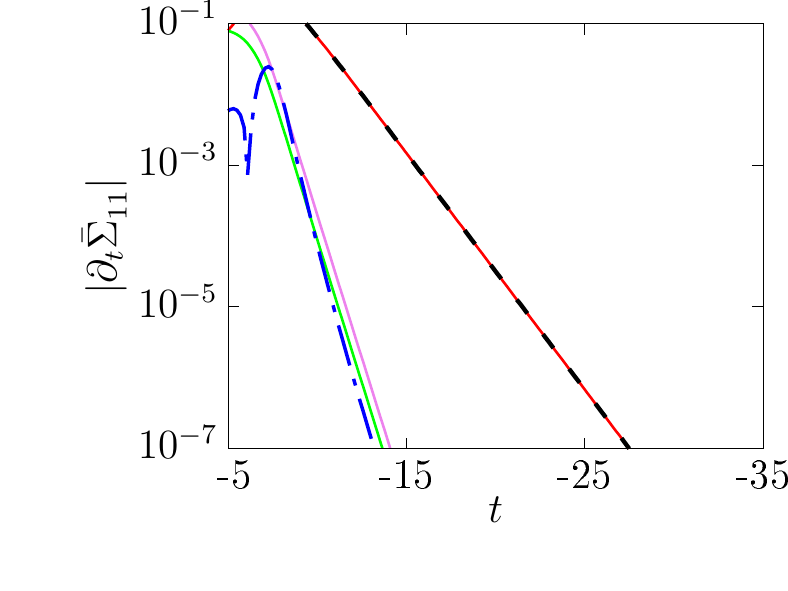} \\
\includegraphics[width=0.3\linewidth]{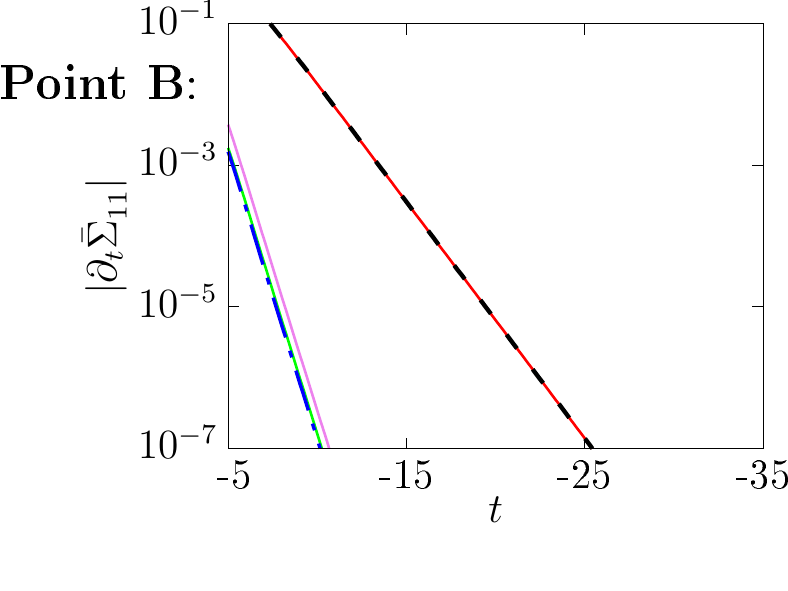}
\includegraphics[width=0.3\linewidth]{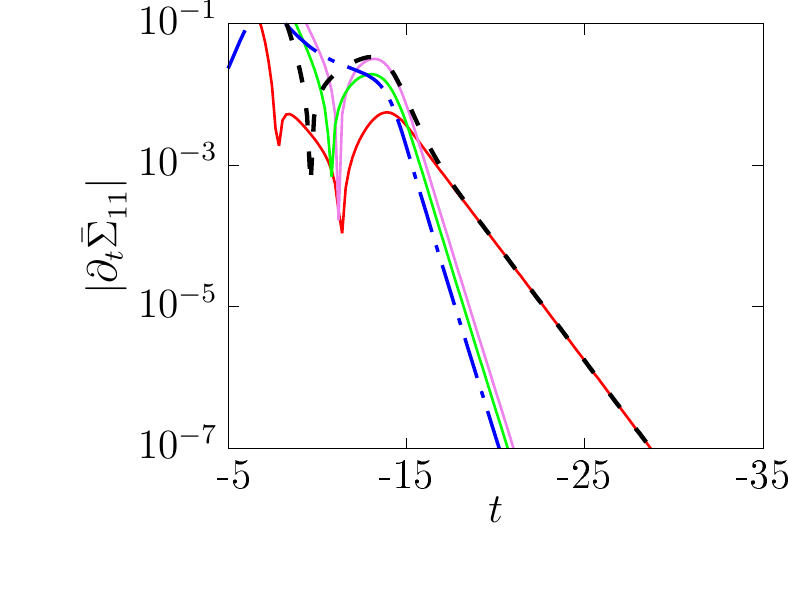}
\includegraphics[width=0.3\linewidth]{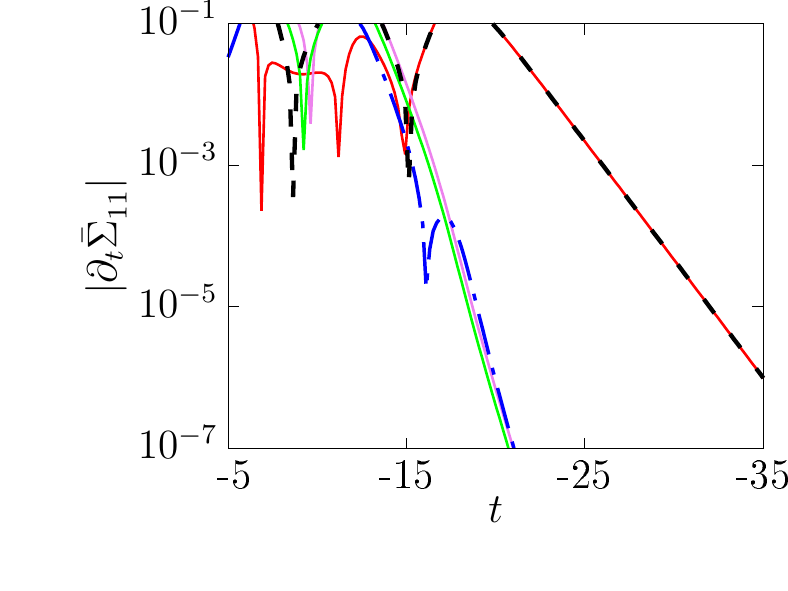} \\
\includegraphics[width=0.3\linewidth]{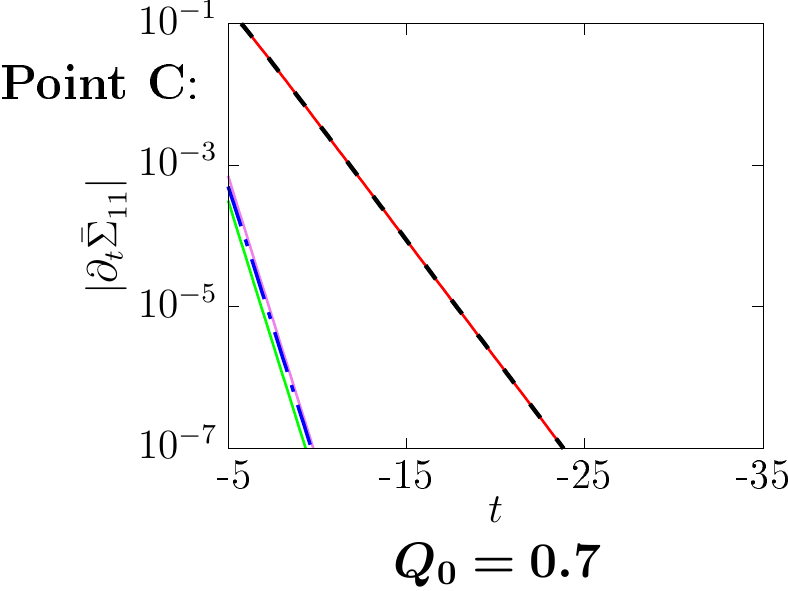}
\includegraphics[width=0.3\linewidth]{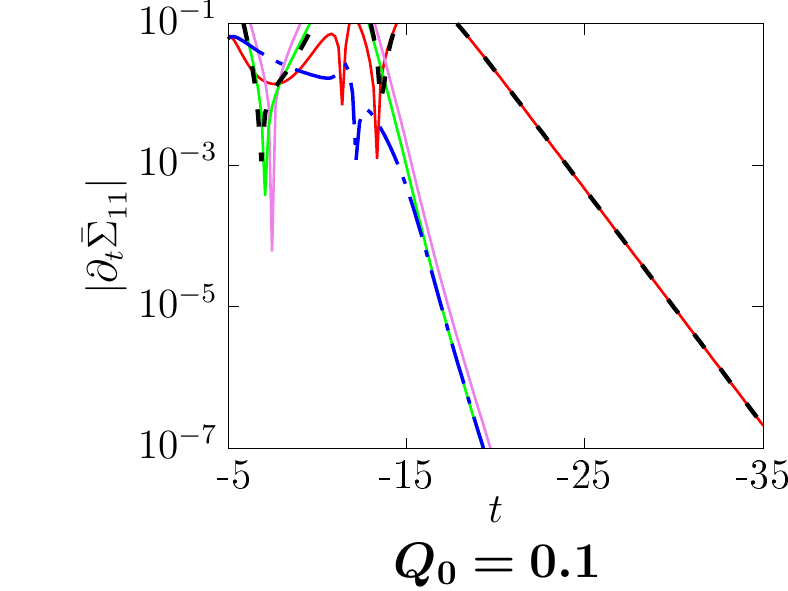}
\includegraphics[width=0.3\linewidth]{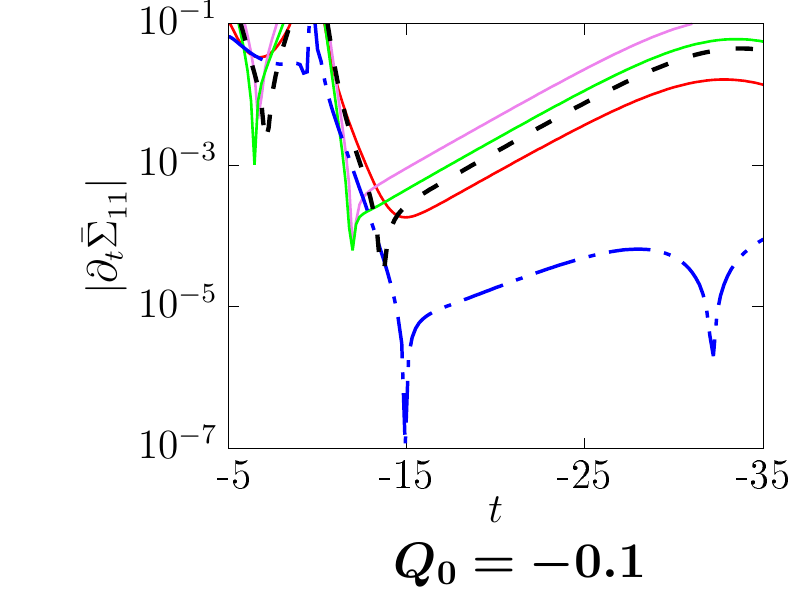} \\
\caption{
Evolution of $|\partial_t \tensor {\bar{\Sigma}}{_1_1}|$: columns correspond to $Q_0=0.7$(left); $Q_0=0.1$ (middle), and $Q_0=-0.1$ (right). Rows correspond to points  $A$ (top), $B$ (middle), and $C$ (bottom), as described in the text.
The different curves correspond to the magnitudes of different contributions, as indicated in the legend, 
to the $(11)$-component of Eq.~\eqref{eq-sigma-ab}.
}
 \label{Figure6}
\end{center}
\end{figure}

\begin{figure}[tb]
\begin{center}
\includegraphics[width=0.3\linewidth]{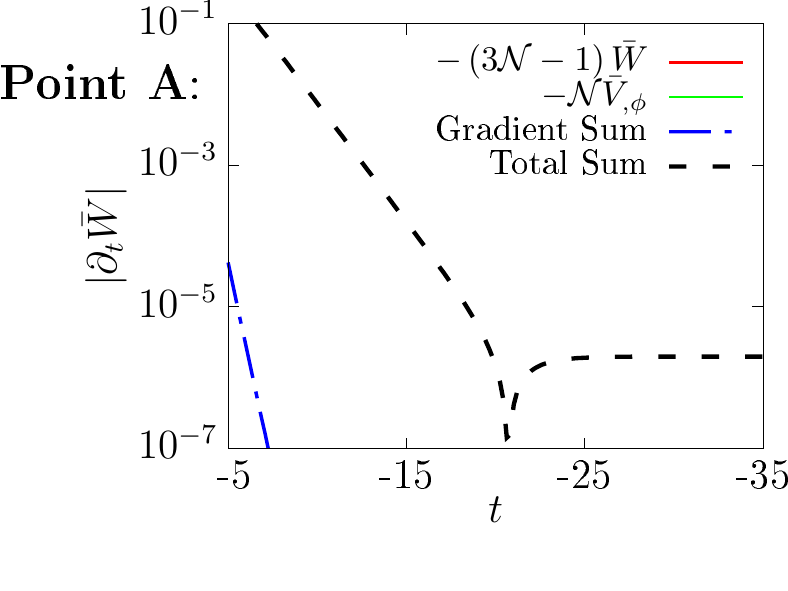}
\includegraphics[width=0.3\linewidth]{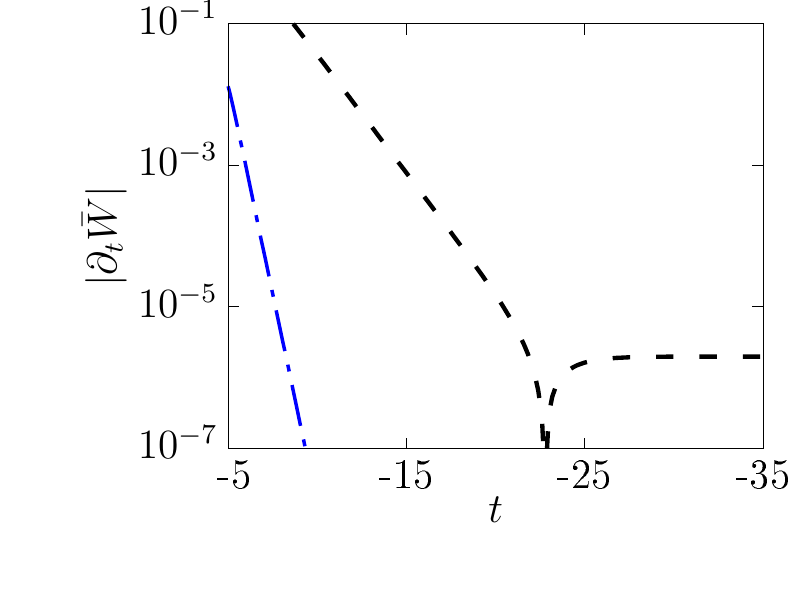}
\includegraphics[width=0.3\linewidth]{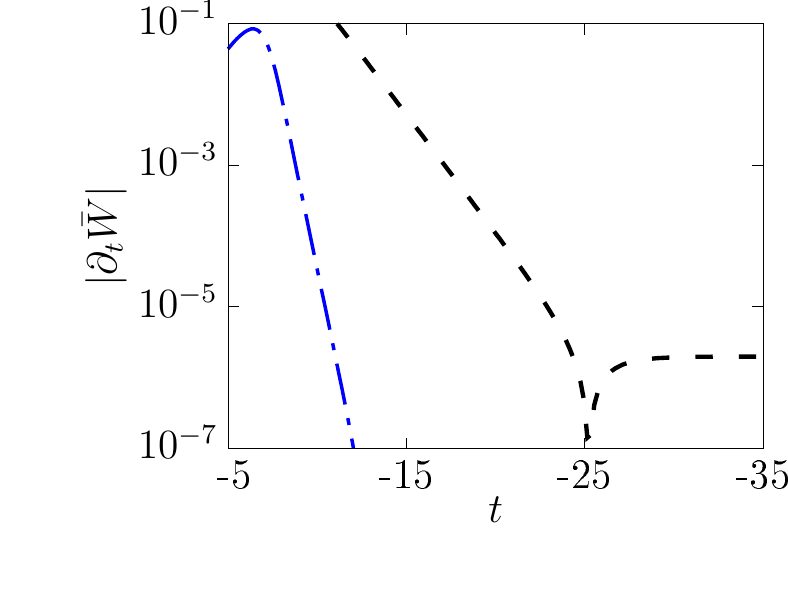} \\
\includegraphics[width=0.3\linewidth]{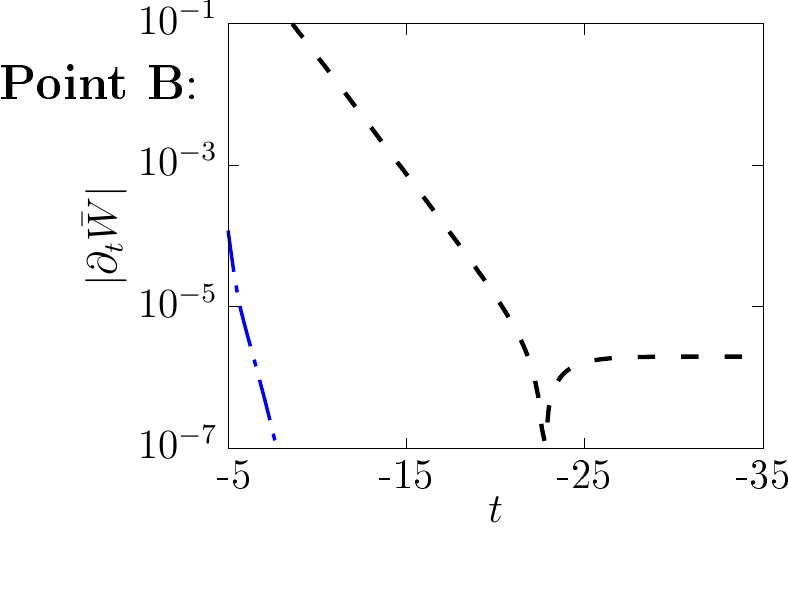}
\includegraphics[width=0.3\linewidth]{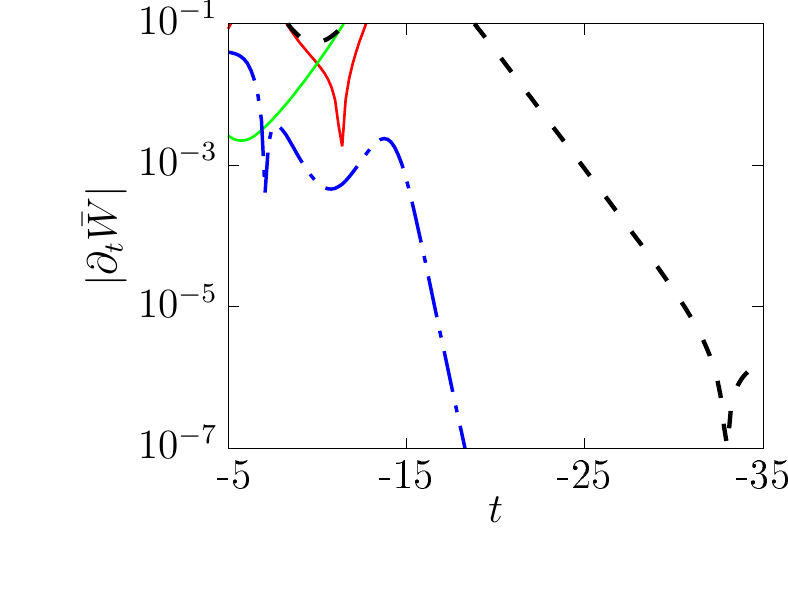}
\includegraphics[width=0.3\linewidth]{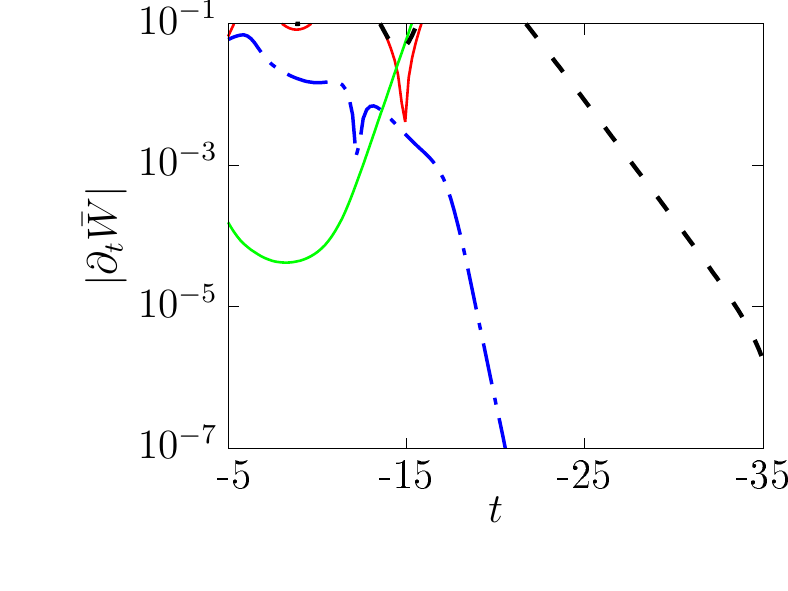} \\
\includegraphics[width=0.3\linewidth]{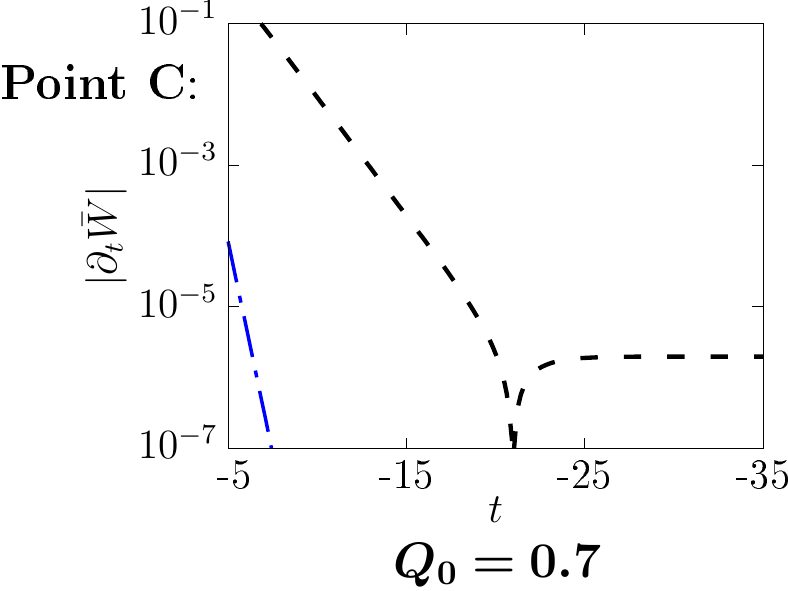}
\includegraphics[width=0.3\linewidth]{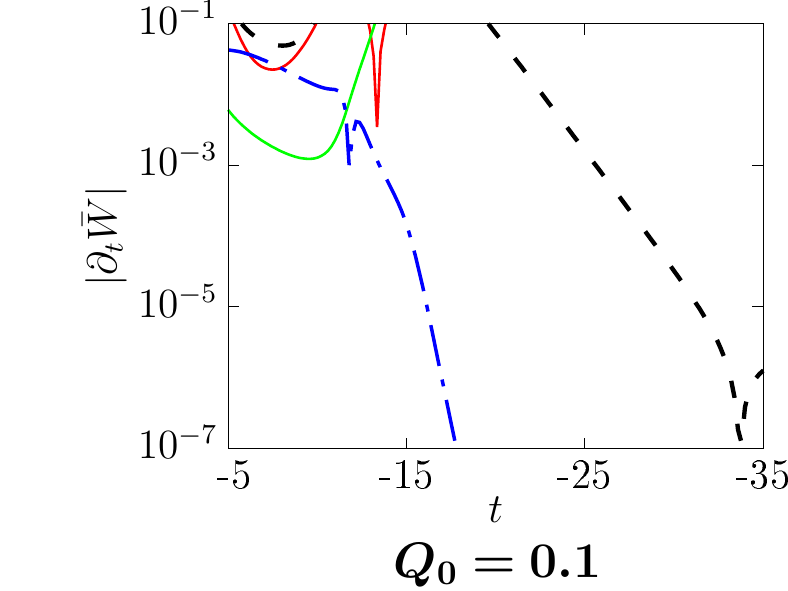}
\includegraphics[width=0.3\linewidth]{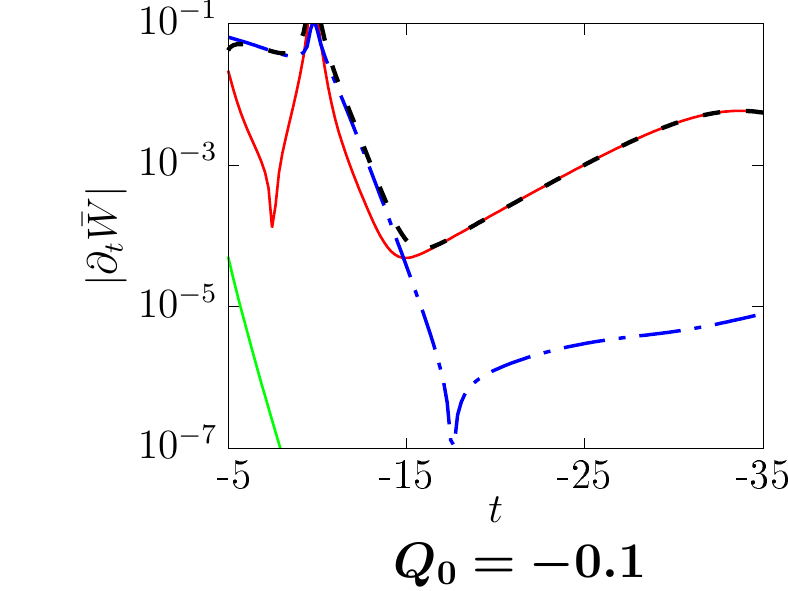} \\
\caption{
Evolution of $|\partial_t \bar{W}|$: columns correspond to $Q_0=0.7$(left); $Q_0=0.1$ (middle), and $Q_0=-0.1$ (right). Rows correspond to points  $A$ (top), $B$ (middle), and $C$ (bottom), as described in the text.
The different curves correspond to the magnitudes of different contributions, as indicated in the legend, 
to the $(11)$-component of Eq.~\eqref{eq-w-Hn}.
The flat-lining at large $t$ observed in some panels is a result of reaching the resolution floor for those terms.
}
 \label{Figure7}
\end{center}
\end{figure}

\begin{figure}[tb]
\begin{center}
\includegraphics[width=0.3\linewidth]{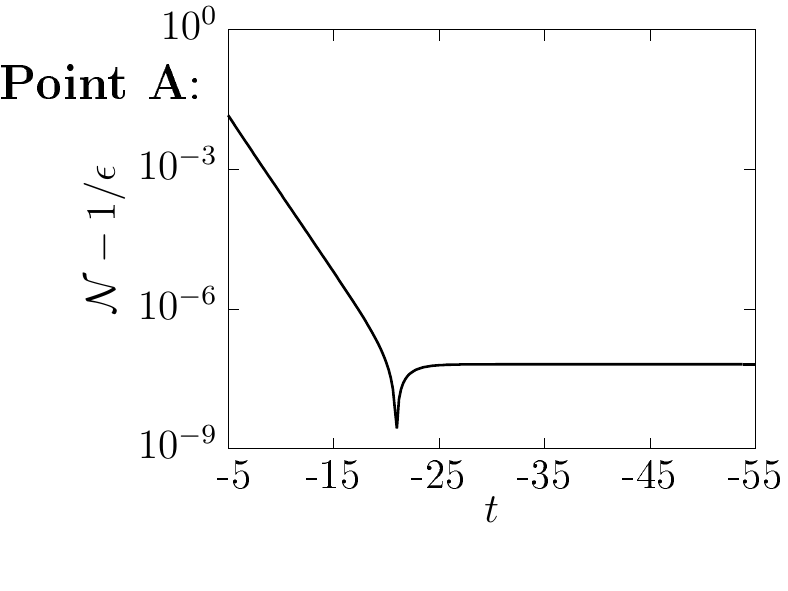}
\includegraphics[width=0.3\linewidth]{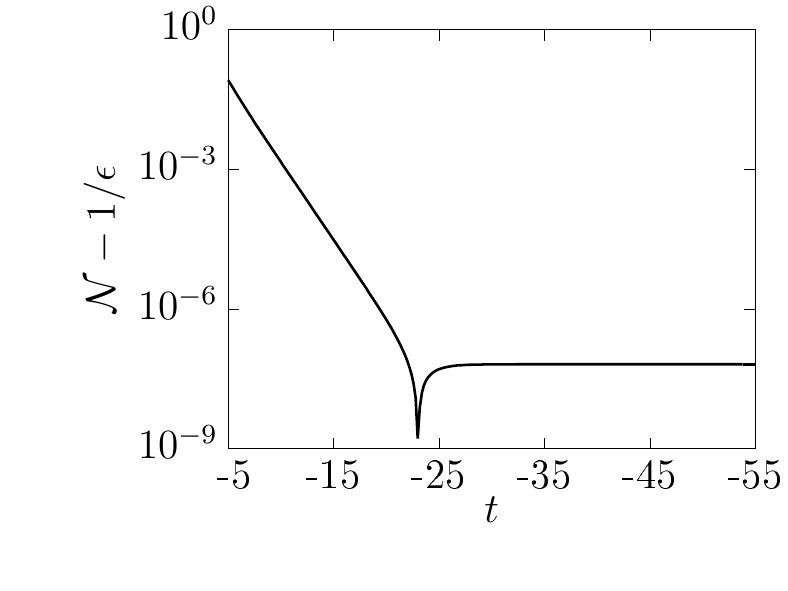}
\includegraphics[width=0.3\linewidth]{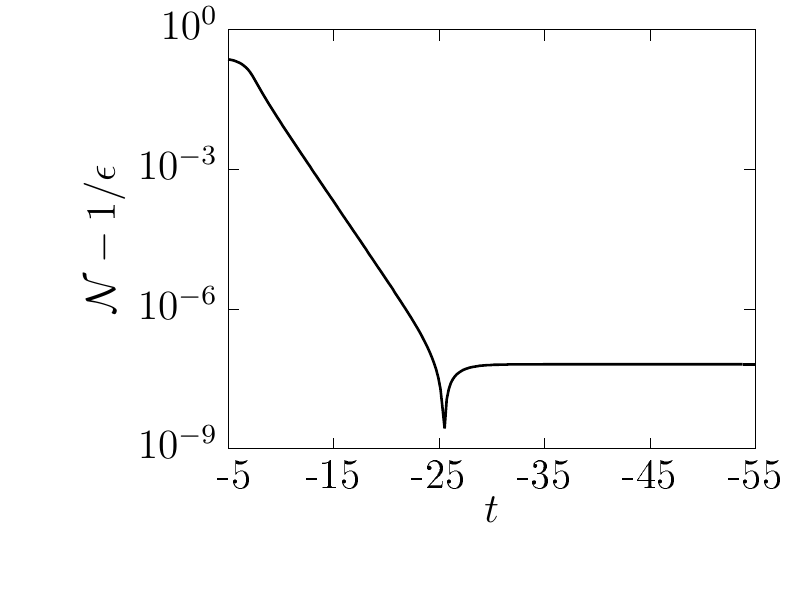} \\
\includegraphics[width=0.3\linewidth]{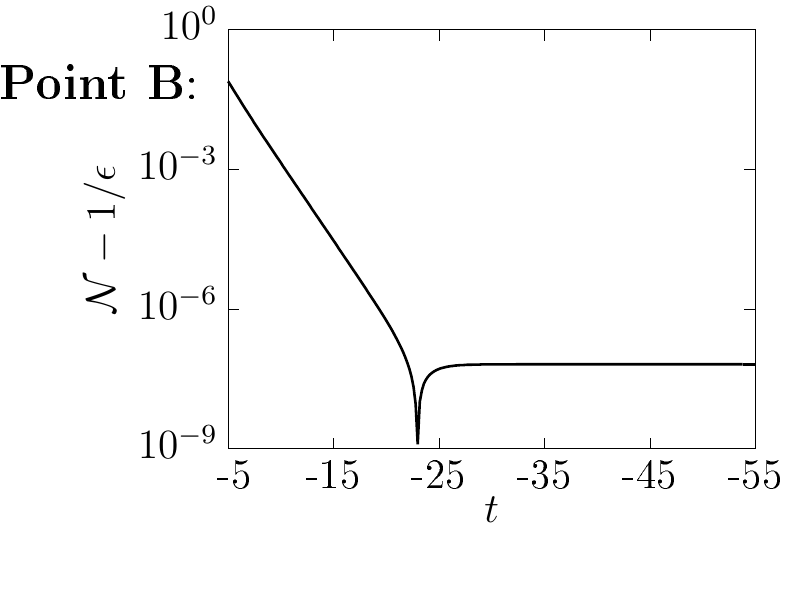}
\includegraphics[width=0.3\linewidth]{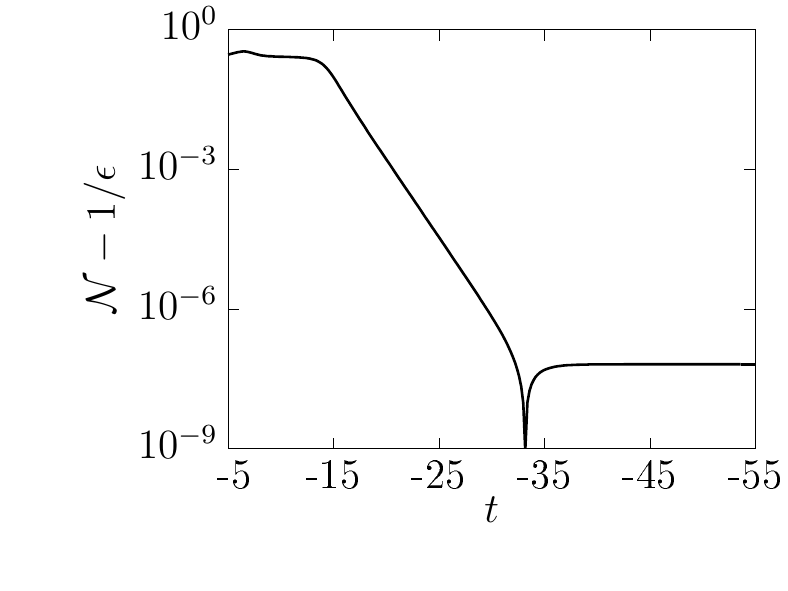}
\includegraphics[width=0.3\linewidth]{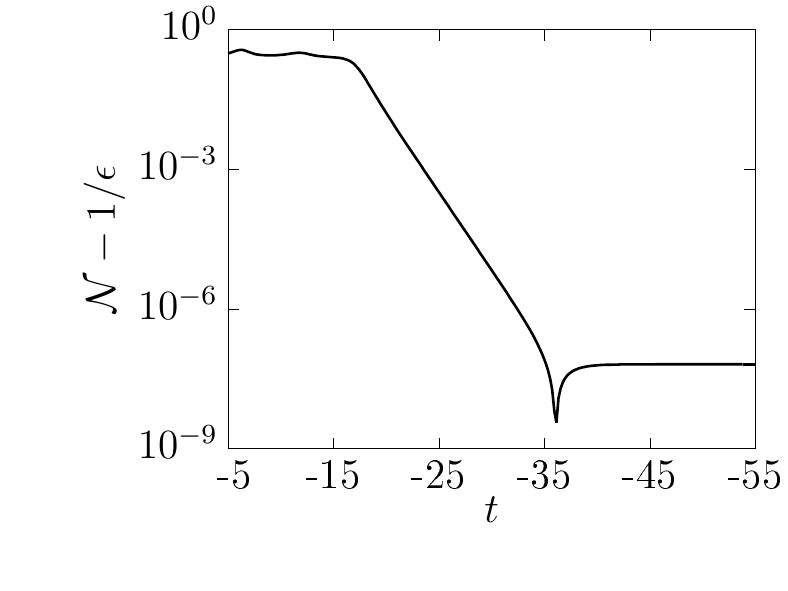} \\
\includegraphics[width=0.3\linewidth]{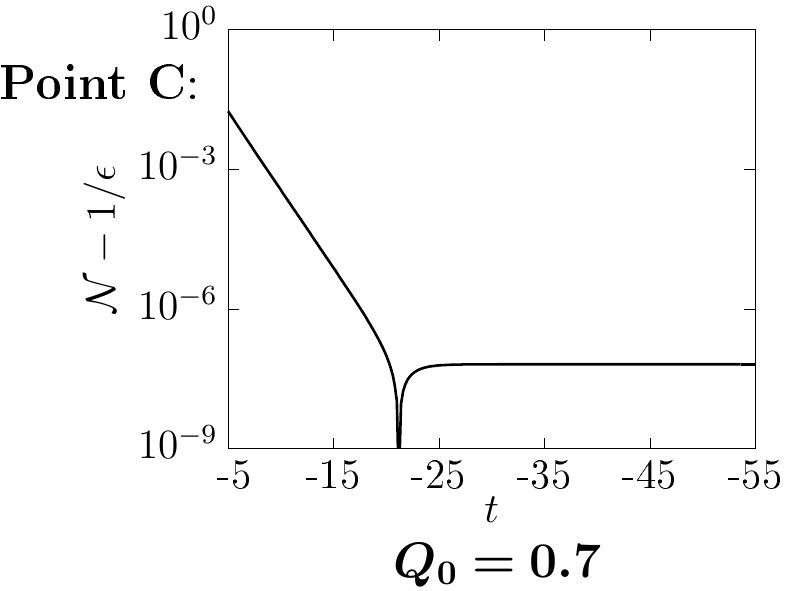}
\includegraphics[width=0.3\linewidth]{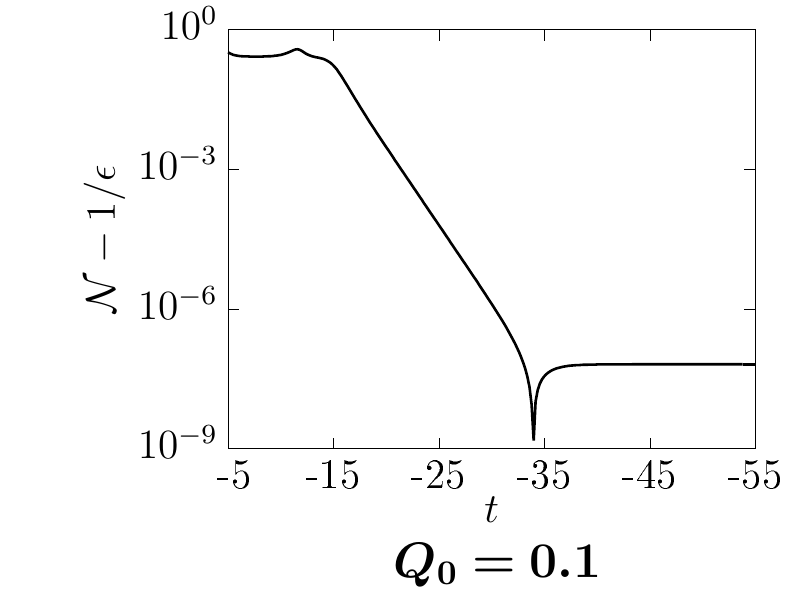}
\includegraphics[width=0.3\linewidth]{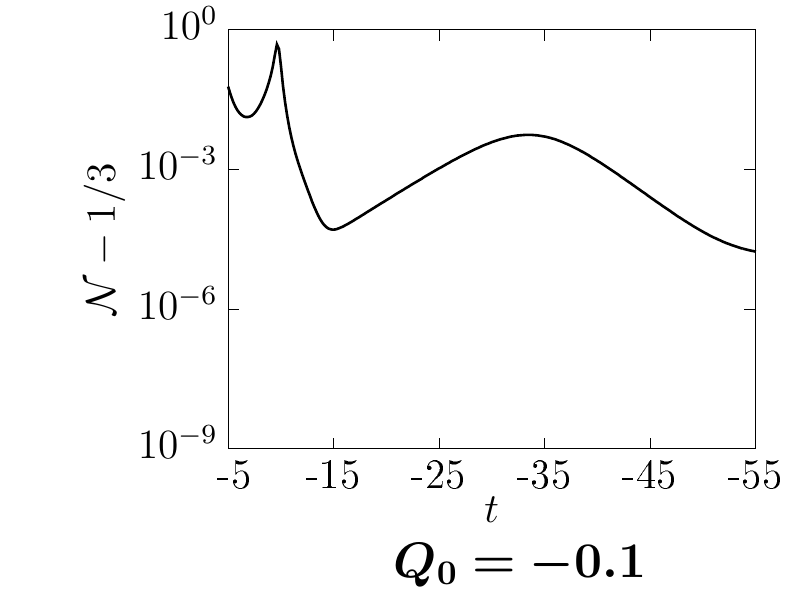} \\
\caption{
Evolution of $|{\cal N} - 1/\varepsilon|$, where $\varepsilon$ is the value at the 
end of the simulation $t_{\rm end}=-150$.  For the first eight panels, $\varepsilon=13$ and 
for bottom-right most panel $\varepsilon=1/3$.
Columns correspond to $Q_0=0.7$(left); $Q_0=0.1$ (middle), and $Q_0=-0.1$ (right). Rows correspond to points  $A$ (top), $B$ (middle), and $C$ (bottom), as described in the text.
The flat-lining at large $t$ observed in some panels is a result of reaching the resolution floor.}
 \label{Figure8}
\end{center}
\end{figure}

Figures~\ref{Figure5}-\ref{Figure8} present the behavior 
$|\partial_t \bar{n}_{11}(t)|$, $|\partial_t \bar{\Sigma}_{11}(t)|$, $|\partial_t \bar{W}(t)|$, and ${\cal N}(t)$, respectively,
 at each of the three lines $A$, $B$ and $C$ (rows)
for the three simulations (columns).   The full simulation runs from $t=0$ to $t_{\rm end}=-150$, but we 
have truncated the $t$-axis and expanded the ordinate axis to expose the detailed features. The geometric variables $\bar{n}_{ab}$ and $\bar{\Sigma}_{ab}$ for 
$(ab) \ne (11)$  display very similar behavior.   For Figs.~\ref{Figure5}-\ref{Figure7}, the behavior of the total 
gradient-dependent contributions and the individual velocity contributions have been plotted in addition to 
the sum over all contributions since these are important for identifying the smoothing stages.

In all but one panel, corresponding to Point $C$ in the $Q_0=-0.1$ simulation (bottom row, rightmost column), smoothing to flat FRW is reached by $t=35$, well before $t_{\rm end}=-150$.  The first feature to note in these cases 
 is that the smoothing occurs through a sequence
of well-defined stages that are the same independent of the local initial conditions, although the stages can occur
at different times.  

During the first stage, the velocity and gradient contributions to the Einstein-scalar field system of equations are roughly comparable and 
may vary up and down ({\it e.g.}, see the case
of Point $B$ in the $Q_0=-0.1$ simulation; bottom row middle panel); the details  depend on the local initial conditions.  In regions of spacetime where the initial velocity
terms begin much larger than the gradient contributions ({\it e.g.},  Point $A$ in the $Q_0=0.7$ simulation; top row left panel) the evolution proceeds directly to the next stage.

During the second stage, the gradient and the sub-leading velocity
terms begin to drop uniformly and  exponentially with time compared to the leading velocity term.  
As this occurs,  the leading
velocity term increasingly dominates and the   evolution 
approaches  {\it ultralocal} behavior; that is, gradient terms become negligible.  
Note that ultralocal -- even ultralocal at every spacetime point in a simulation -- is not equivalent to
flat FRW because, in principle, all velocity  terms can be large and there can remain long wavelength variations 
in spatial curvature and shear.  
This stage ends when the magnitudes of the gradient and subleading velocity terms have decreased sufficiently. As our condition, we choose a decrease by six orders of magnitude. (This choice is somewhat arbitrary. For example, we could have chosen four orders of magnitude. We adopt this convention based on the fact that in our simulations the resolution floor for certain terms is $\approx 10^{-6}$.)
During the same stage, the Hubble-normalized lapse 
${\cal N}(t)$ approaches the value it maintains for the remainder of the simulation, ${\cal N}=1/\varepsilon$.
As can be seen from Figs.~\ref{Figure5}-\ref{Figure8}, the duration of the second stage is nearly the same 
independent of initial conditions, although the start and end times may differ.

The third stage consists of purely ultralocal evolution in which the remaining (leading) velocity terms 
decrease uniformly exponentially, though at a slower rate than the subleading velocity and gradient terms did. 
The third stage ends when the leading velocity terms reach zero (to within the resolution limit of the simulation).  
As was observed for the second stage, the duration of the third stage is nearly the same 
for all regions, although the start and end times may differ. 

The final stage is the flat FRW attractor stage that endures for the remainder of the simulation.
 
The nearly-universal behavior during the second and third stages can be understood 
 by first considering only the leading velocity contributions to
Eqs.~(\ref{eq-n-ab}), (\ref{eq-sigma-ab}), and Eq.~(\ref{eq-w-Hn}) that come to dominate
during the second stage:
\begin{eqnarray}
\label{eq-n-ab2}
\partial _t \bar{n}_{ab} &\approx& - \Big({\cal N} - 1 \Big) \bar{n}_{ab}
,\\
\label{eq-sigma-ab2}
\partial _t \bar{\Sigma}_{ab} &\approx& - \Big( 3 {\cal N} - 1 \Big) \bar{\Sigma}_{ab}
,\\
\label{eq-w-Hn2}
\partial_t \bar{W} &\approx & - \Big(  3 {\cal N} -1 \Big) \bar{W} 
- {\cal N} \bar{V}_{,\phi}.
\end{eqnarray}
The solutions to the first equations are 
\begin{equation} {\rm ln } \, \bar{n}_{ab}=  -({\cal N} - 1) \, t, \quad  {\rm ln}  \, \bar{\Sigma}_{ab}=  -(3 {\cal N} - 1 ) \, t.
\end{equation}
These expressions, 
combined with the fact that ${\cal N}\approx 1/\varepsilon$, accurately explain the uniform 
slopes on the log-plots observed during the third (purely ultralocal evolution) stage when  
these leading terms are the only non-zero contributions.  (Recall from 
Eq.~\eqref{time-choice} that the time coordinate has been chosen so that its sign
 is negative and $t$ runs towards $-\infty$ as slow contraction proceeds.)  

Finally, it is straightforward to recast the flat FRW attractor solution as given in Eq.~\eqref{FRW-scaling-sol} as 
\begin{equation}
\bar{W} =- \sqrt{2 \varepsilon}\quad {\rm and} \quad
\bar{V}_{,\phi} = \sqrt{2 \varepsilon}\, (3 -\varepsilon).
\end{equation} 
In particular, the two velocity
contributions to $\partial_t \bar{W}$ in Eq.~\eqref{eq-w-Hn2} are equal and opposite.  Their difference decreases exponentially during the 
 the ultralocal phase in proportion to ${\cal N} -1/\varepsilon$, with both approaching
 zero (to within the resolution) at the end of the ultralocal stage and the beginning of the final flat FRW attractor stage.

These results for the leading behavior provide an estimate of how the subleading velocity terms behave during 
the second stage just by counting the powers of $\bar{n}_{ab}$ and $\bar{\Sigma}_{ab}$ involved. 
 For example, the subleading velocity contribution to   $\partial_t \bar{n}_{ab}$ in
 Eq.~(\ref{eq-n-ab}), ${\cal N} ( 2  \bar{n}_{(a}{}^c \bar{\Sigma}_{b)c})$ scales with time approximately as the 
 product of $\bar{n}_{ab}$ and $\bar{\Sigma}_{ab}$; or, equivalently, its logarithm is
proportional to $ {\rm ln } \, \bar{n}_{ab} +  {\rm ln } \, \bar{\Sigma}_{ab}= -(4 {\cal N}-2) \, t$, an estimate that is
 in excellent agreement with the simulation results. Similar scaling arguments explain the behavior of
 the gradient terms as well.  This is different from the naive expectation that homogeneous spatial curvature
 ($\propto -k/a^2$) decreases faster than homogeneous Kasner-like anisotropy ($\propto 1/a^6$) during 
 contraction; see Ref.~\cite{Erickson:2003zm}.

For completeness, we comment on the evolution at Point $C$ in the $Q_0 =-0.1$ simulation where smoothing is not 
 reached by the end of the simulation.  It is interesting to note that ${\cal N} \rightarrow 1/3$ rather than
 the smoothing attractor solution  ${\cal N} \rightarrow 1/\varepsilon$, consistent with a phase in which
 the scalar field potential energy is negligible and scalar field energy density is purely kinetic and gradient
 dominated.  This is reminiscent of what is expected in Kasner-like evolution with a free scalar field.  
 However, we would caution that this interpretation is not trustworthy because, as noted in the Appendix, the evolution at and  near Point $C$ exhibit substantial deviations from numerical convergence. Further studies
 are needed to determine what occurs in these regions.  Since they only occur in a very limited range of phase space
 that is atypical for bouncing or cyclic models, we set this aside for future investigation.

\section{Discussion}
\label{sec_discussion} 

In developing improved numerical relativity tools that enable the extension of cosmological simulations
to spacetimes with spatial variations along two independent directions, we have been able to confirm
the robust smoothing effect of slow contraction over a wide range of initial conditions that 
are far from flat FRW (Fig.~\ref{Figure1}), as  found previously using 
codes that only allowed variations along one spatial direction.  

More significantly, it has been possible for the first time to demonstrate two features of the smoothing process.
First, smoothing by slow contraction  obeys to good approximation a
universal behavior that is independent of local initial conditions.  Comparing the examples in 
Figs.~\ref{Figure5}-\ref{Figure8}, the sequence of stages is the same and duration of stages two
and three is the same for any local regions where smoothing occurs.  This statement applies 
for any two regions in the same simulation or for two regions in simulations with different 
initial conditions.  It is also true whether the spacetime is completely smoothed with no 
evidence of spiking or whether the spacetime does not smooth completely.  

Secondly, the work establishes that the common heuristic  explanation of smoothing to flat FRW in the literature,
whether through inflationary expansion or slow contraction, does not capture the dynamics accurately, at least for the case of slow contraction.  
The view has been that ultralocality and homogeneity are reached over a given region of spacetime that lies 
within a single Hubble patch of radius $\Theta$, and then inflation or contraction transforms that homogeneous region into one that spans exponentially many independent Hubble patches.  Once ultralocality and homogeneity within the initial Hubble patch 
are assumed, it is straightforward to show using the Friedmann equations that homogeneous spatial curvature and anisotropy are suppressed across the many final Hubble patches at the end.   

Utilizing the advances in numerical relativity discussed in this paper,  we have been able to show that smoothing 
by slow contraction does not follow this picture.  Instead, beginning with a region that is far from ultralocality and homogeneity and lies within a single Hubble patch, the Hubble radius shrinks to a size much smaller than the region before ultralocality is reached.  By the time the evolution converges to ultralocality (end of stage 2 as described above in Sec.~5) and the next stage begins, the region already spans exponentially many independent Hubble-sized patches.  From there, after a predictable period, the exponentially many 
independent Hubble patches (except perhaps for regions with negligible small physical volume \cite{Lim:2009dg,Ijjas:2020dws}) are each driven towards a flat FRW geometry and together transform the original region into  homogeneous, isotropic and flat regions of spacetime.

Studies of these phenomena  for cases of slow contraction in which the scalar field component 
has less pressure (smaller $\varepsilon$) that is closer
to the critical value $\varepsilon=3$,  for ordinary contraction  $\varepsilon<3$ and for inflationary 
expansion will be the subject of forthcoming publications.

\section*{Acknowledgments}
The work of A.I. is supported by the Lise Meitner Excellence Program of the Max Planck Society and by the Simons Foundation grant number 663083.
A.P.S. and W.G.C. are partially supported by the Simons Foundation grant number 654561. 
F.P. acknowledges support from NSF grant PHY-1912171, the Simons Foundation, and the Canadian Institute For Advanced Research (CIFAR).  P.J.S. is supported in part by the DOE grant number DEFG02-91ER40671 and by the Simons Foundation grant number 654561.  

\newpage
\appendix

\section{Numerical methods and convergence}
\label{sec_convergence}

In this Appendix, we describe our tests for numerical convergence.  The key results are:
\begin{itemize}
\item For initial data for which the evolution leads to smoothing everywhere by $t_{\rm end}$ without spiking, the simulation strongly satisfies all convergence tests at all spacetime points, as represented by the case with average initial scalar field velocity $Q_0 =0.7$.  This represents the overwhelming majority of the phase diagram shown in Fig.~\ref{Figure1}.
\item  For initial data for which the evolution leads to smoothing everywhere by $t_{\rm end}$ but with spiking in some local
regions, the simulation strongly satisfies all convergence tests at all spacetime points outside the spiking regions, as represented by the case with average initial scalar field velocity $Q_0 =0.1$.  
\item  For initial data for which the evolution does not smooth in some local regions by $t_{\rm end}$, it can still be that most of the volume that does smooth strongly satisfies all convergence tests, as represented by the case with average initial scalar field velocity $Q_0 =-0.1$.  
\end{itemize}

\begin{figure}[b]
\begin{center}
{\includegraphics[width=6.25in,angle=-0]{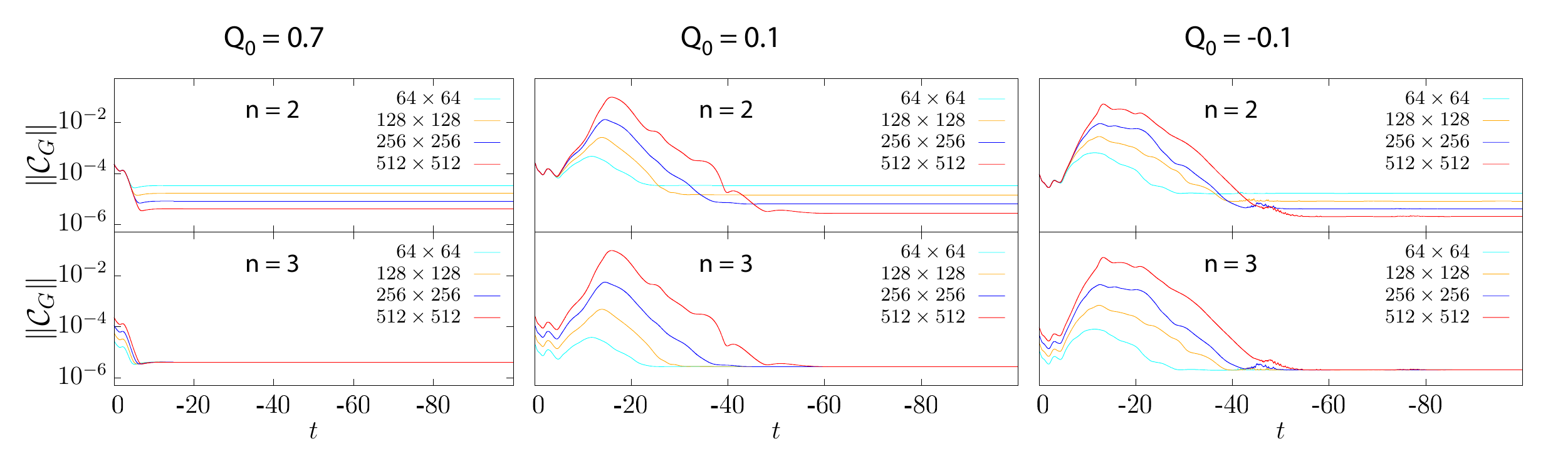}}
\caption{
Rescaled L2 norm of the Hamiltonian constraint $\Vert {\cal C}_{\rm G} \Vert$  for four resolutions.   Columns correspond to $Q_0=0.7$ (left), $Q_0=0.1$ (middle), and $Q_0=-0.1$ (right). The highest resolution spacetime grid is $512 \times 512$ in each case. To check for the order of convergence, we repeat the simulation with lower-resolution $D \times D$ grids with $D = \{256, 128, 64\}$ and divide 
$\Vert {\cal C}_{\rm G} \Vert$ for these resolutions by $(512/D)^n$ where $n=2$ for the second order convergence test and $n=3$ for the third order test.  In the case $Q=0.7$, the simulation runs directly from second order to third order convergence (top and bottom panels in left column). For the other cases, there is a time interval of non-convergence in between. 
}
\label{figapp:conv}
\end{center}
\end{figure}

To numerically solve the system of equations detailed in Eqs.~(\ref{eq-E-ai-Hn}-\ref{eq-barS-Hn}, \ref{Neqn}), we use second order accurate spatial derivatives, and a three step method for time integration given by the Iterated Crank-Nicolson method. The evolution equations consist of a coupled elliptic-hyperbolic system of equations, so at each sub-step of the time integration, we first solve the elliptic equation for the Hubble-normalized lapse $\mathcal{N}$ using a multigrid V-cycle method with six subgrids and then update the hyperbolic equations to the next Iterated Crank-Nicolson sub-step. In the simulations illustrated above we use a $256 \times 256$ grid, with $ \Delta x = \Delta y = 2\pi/256$ and a Courant factor of $0.5$.  We have also computed results for $64 \times 64$, $128 \times 128$ and $512 \times 512$ grids in conducting our convergence tests, as described below.

Fig.~\ref{figapp:conv} shows  the L2 norm of the Hamiltonian constraint, ${\cal C}_{\rm G}=0$, as given in Eq.~\eqref{constraintG} integrated over the spatial domain as a function of time, where
\begin{equation}
{\cal C}_{\rm G} \equiv
3 + 2 \bar{E}_a{}^i \partial _a \bar{A}^a - 3 \bar{A}^a \bar{A}_a
- {\textstyle \frac12 } \bar{n}^{ab} \bar{n}_{ab}
+ {\textstyle \frac14 } ( \bar{n}^c{}_c)^2 
- {\textstyle \frac12 } \bar{\Sigma}^{ab} \bar{\Sigma}_{ab}
- {\textstyle \frac12 } \bar{W}^2 -  {\textstyle \frac12 } \bar{S}^a \bar{S}_a -   {\bar V} 
\,.
\end{equation}
\begin{figure}[b]
\begin{center}
\includegraphics[width=0.3\linewidth]{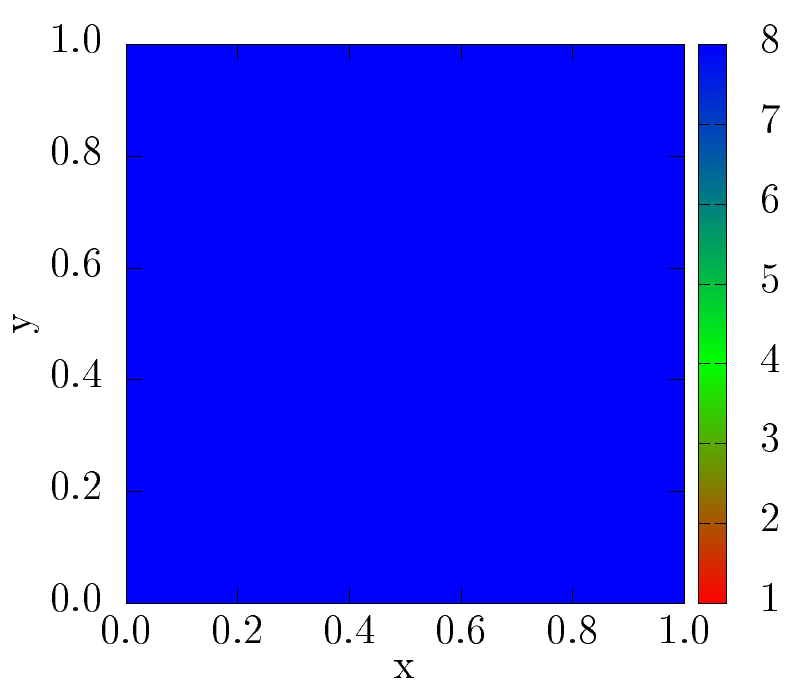}
\includegraphics[width=0.3\linewidth]{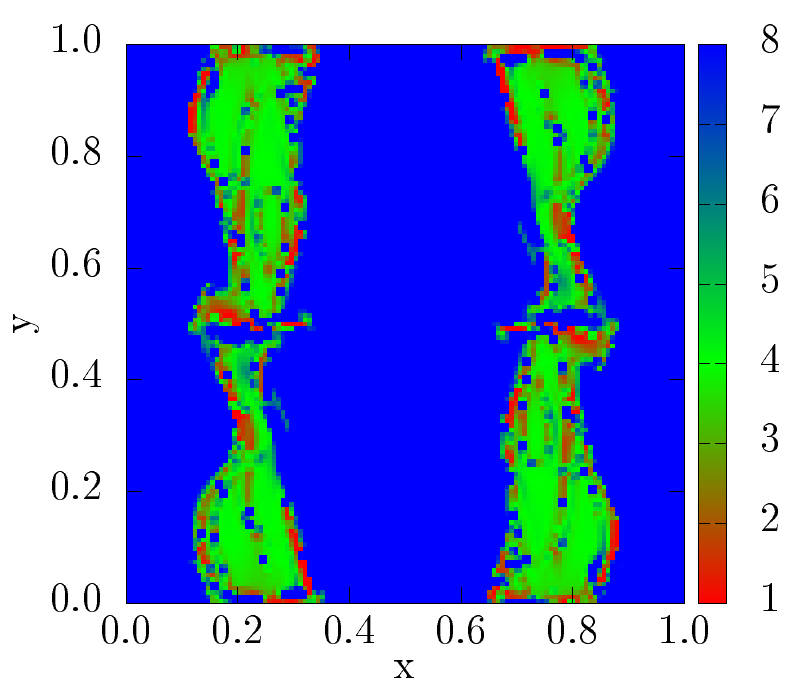}
\includegraphics[width=0.3\linewidth]{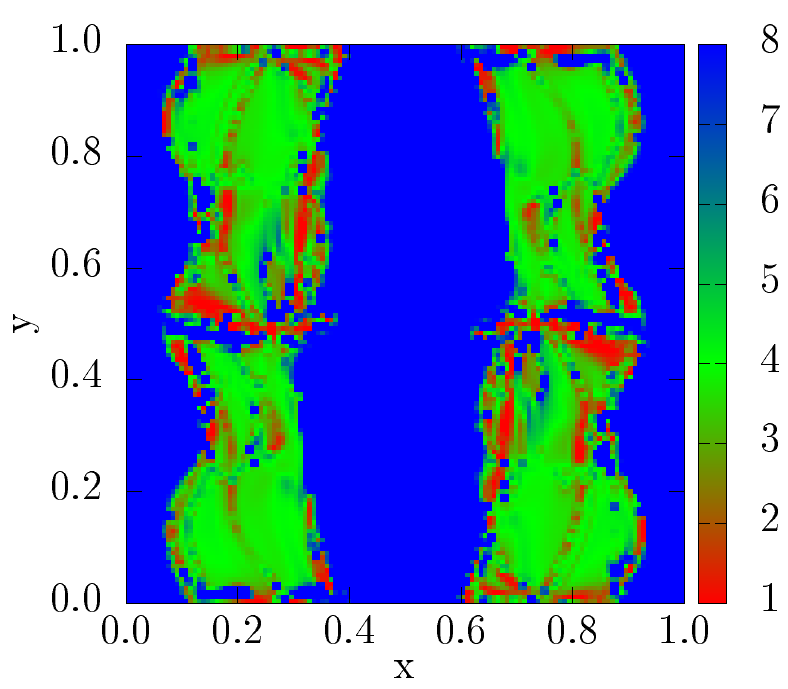} \\
\includegraphics[width=0.3\linewidth]{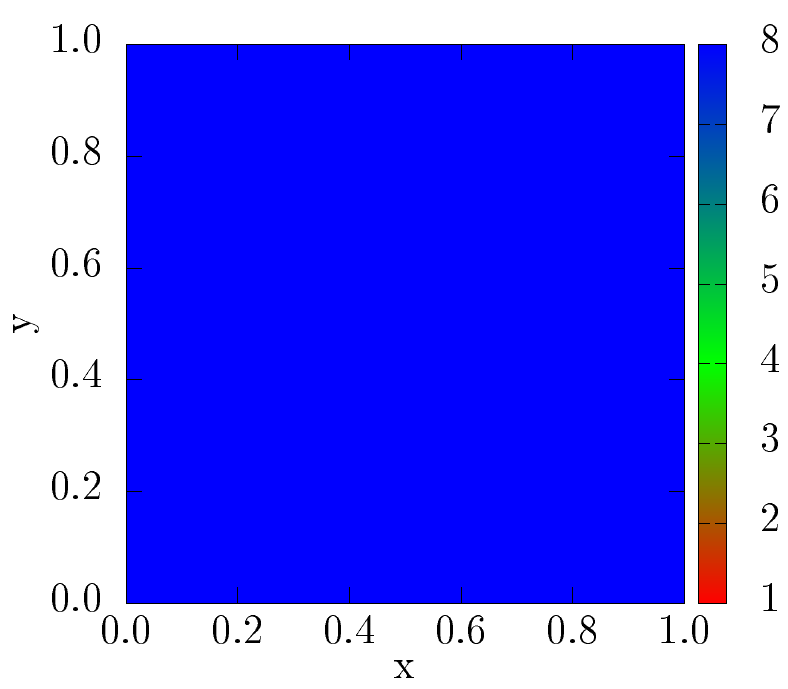}
\includegraphics[width=0.3\linewidth]{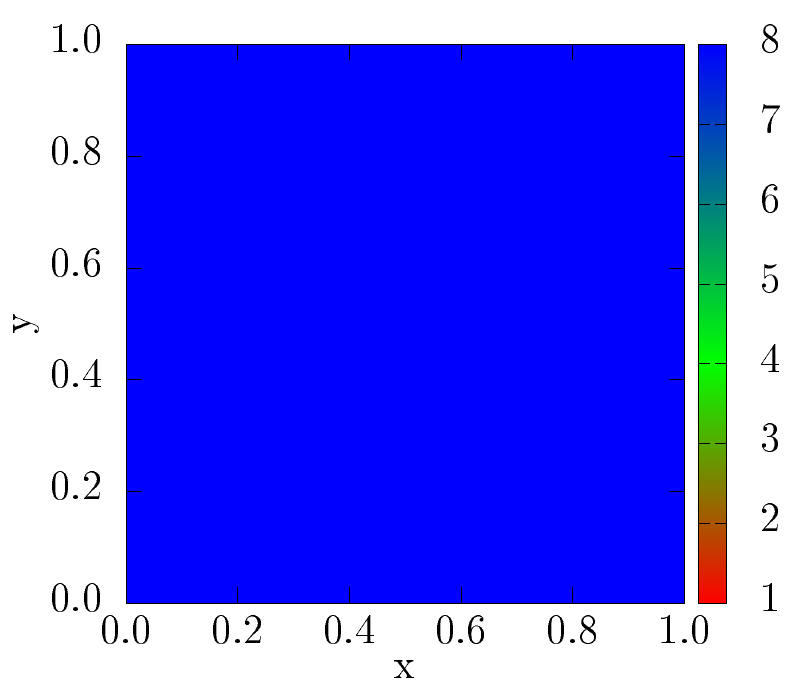}
\includegraphics[width=0.3\linewidth]{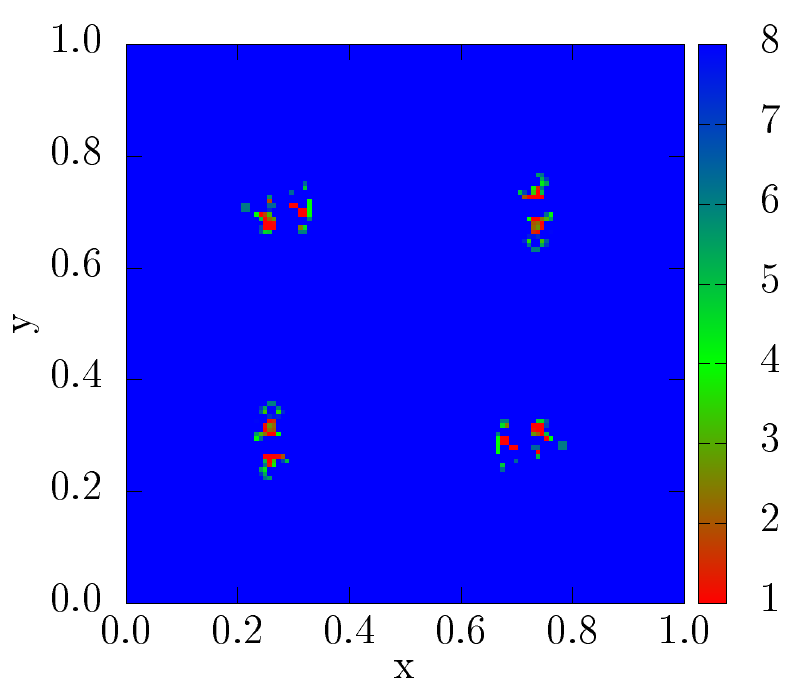}
\caption{
Top view map showing regions of second order (green) and third order (blue) convergence and regions of non-convergence (red) based on 
comparing simulations with $128 \times 128$ and $256 \times 256$ grid resolutions.  
Three models with $Q_0={0.7,0.1,-0.1}$ (left, middle and right columns) are compared
 at $t_{\rm spike}=-15$ (top row), the time when the L2 norm exhibits the maximum deviation 
 from convergence,  and at the end of the simulation $t_{\rm end}= -150$ (bottom row).}
\label{figapp:cgmap}
\end{center}
\end{figure}
 The norm is computed 
 for four different resolutions rescaled by the convergence factor for the three cases of $Q_0=0.7$ (left), $Q_0=0.1$ (middle), and $Q_0=-0.1$ (right). 
 We see that, in the case of a spacetime that smooths directly to FRW everywhere ($Q_0=0.7$), after an initial period of second order convergence, the evolution moves to directly third order convergence, hence convergent for all $t$.  
For  both the $Q_0=0.1$ and $Q_0=-0.1$ cases, the L2 norm first converges to second order for a period, then loses convergence for a period with a maximum loss of convergence occurring at $t \approx -15$. By $t \approx -50$ both the $Q_0=0.1$ and $Q_0=-0.1$ cases recover convergence and reach third order convergence for the rest of the simulation.

We note that the L2 norm measures the average over the entire spacetime and so, by itself, does not allow us to discriminate regions that strongly satisfy all convergence tests at all times from those that do not.  To distinguish these, we have constructed maps 
exploring the convergence factor of our entire domain over time, as shown in Fig.~\ref{figapp:cgmap}.
The maps demonstrate that the regions of lost convergence are comparably small (exponentially small when normalized by 
conformal volume)  and correlate with regions where spikes develop.

\newpage

\bibliographystyle{plain}
\bibliography{bib_long_paper}

\end{document}